\documentclass[twocolumn,showpacs,preprintnumbers,superscriptaddress,amsmath,amssymb,floatfix,nofootinbib,pra]{revtex4-1}

\usepackage{graphicx}
\usepackage{bm}
\usepackage{subfigure,bbold}
\usepackage{color}
\usepackage[normalem]{ulem}

\newcommand{\mri}{\mathrm{i}}
\newcommand{\tr}{\mbox{Tr}}

\newcommand{\Exp}[1]{\mathrm{e}^{\mbox{\footnotesize$#1$}}}
\renewcommand{\vec}[1]{\bm{#1}}
\newcommand{\bs}[1]{{\boldsymbol#1}}

\newcounter{query}
\begin{document}
\title{Analytical and numerical study of uncorrelated disorder on a honeycomb
lattice}

\author{Kean Loon \surname{Lee}}
\affiliation{Centre for Quantum Technologies, National University of Singapore,
3 Science Drive 2, Singapore 117543, Singapore}

\author{Beno\^it \surname{Gr\'emaud}}
\affiliation{Laboratoire Kastler-Brossel, UPMC-Paris 6, ENS, CNRS; 4 Place
Jussieu, F-75005 Paris, France}
\affiliation{Centre for Quantum Technologies, %
National University of Singapore, 3 Science Drive 2, Singapore 117543, %
Singapore}
\affiliation{Department of Physics, National University of Singapore, %
2 Science Drive 3, Singapore 117542, Singapore}

\author{Christian \surname{Miniatura}}
\affiliation{INLN, Universit\'e de Nice-Sophia Antipolis, CNRS; 1361 route
des Lucioles, 06560 Valbonne, France}
\affiliation{Centre for Quantum Technologies, %
National University of Singapore, 3 Science Drive 2, Singapore 117543, %
Singapore}
\affiliation{Department of Physics, National University of Singapore, %
2 Science Drive 3, Singapore 117542, Singapore}

\author{Dominique \surname{Delande}}
\affiliation{Laboratoire Kastler-Brossel, UPMC-Paris 6, ENS, CNRS; 4 Place
Jussieu, F-75005 Paris, France}

\begin{abstract}
We consider a tight-binding model on the regular honeycomb lattice with
uncorrelated on-site disorder. We use two independent methods (recursive
Green's function and self-consistent Born approximation) to extract the scattering mean free path, the scattering
mean free time, the density of states and the localization length as a function
of the disorder strength. The two methods give excellent quantitative
agreement for these single-particle properties. Furthermore, a finite-size
scaling analysis reveals that all localization lengths for different lattice
sizes and different energies (including the energy at the Dirac points)
collapse onto a single curve, in agreement with the one-parameter scaling
theory of localization. The predictions of the self-consistent theory of
localization however fail to quantitatively reproduce these numerically-extracted localization
lengths.
\end{abstract}

\begin{widetext}
\maketitle
\end{widetext}

\section{Introduction}
Anderson localization (AL) of waves~\cite{anderson1958} in disordered
media is a ubiquitous phenomenon which has been observed both for
classical and quantum waves, e.g.  light~\cite{albada1985,bidel2002},
acoustics~\cite{hu2008,faez2009}, water waves~\cite{belzons1988},
ultracold atoms~\cite{billy2008,lemarie2008}, polaritons~\cite{cheng1990}
and quantum Hall system~\cite{ilani2004}. The scaling theory of
localization~\cite{abrahams1979} predicts that a three-dimensional (3D)
system exhibits a metal-insulator transition while 1D and 2D systems
always display localization at any finite disorder strength. Approximate analytical
expressions for the localization length in terms of the transport mean free
path can be derived within the framework of the self-consistent theory of
localization~\cite{vollhardt1980,*wolfle2010,kuhn2007}. Dimension two is
in fact the critical dimension for AL and symmetry considerations can play
an important role. Indeed, while localization is expected to take place for
spinless time-reversal invariant systems (albeit with an exponentially large
localization length), perturbative renormalization group studies
on non-linear $\sigma$-models suggest that a metal-insulator transition may
occur in 2D if chirality is present~\cite{ostrovsky2006,mirlin2010}. Such
a disordered system with different chiral classes could be realized with
the honeycomb lattice. The successful isolation of graphene flakes in
2004~\cite{novoselov2004}, and the discovery that graphene samples exhibit a
finite electronic conductivity at half-filling although the density of states
(DoS) vanishes~\cite{zhang2005,novoselov2005}, has thus spurred interest
in studying electronic transport in graphene in the presence of disorder,
see~\cite{mucciolo2010} and references therein.

However, even though graphene is a readily-available physical realization
of a honeycomb lattice,
its properties are invariably affected by the combined effects of interaction,
disorder and phonons.
The controlled study of disorder alone in graphene sheets
is thus difficult, notwithstanding the fact that engineering
disorder with given statistical properties seems out of reach. In
that respect, ultracold atoms loaded on a graphene-like optical
lattice~\cite{lee2009,sengstock2011, tarruell2012} offer an alternative
route and have already proven their key impact in weak and strong localization
studies~\cite{labeyriel2003_1,*labeyriel2003_2,clement2005,inguscio2005,billy2008,jendrzejewski2012,greiner2002}.
Furthermore, key transport quantities, like the scattering and transport
mean free paths or the localization length, have already been analyzed for
speckle optical potentials in the Born approximation in Ref.~\cite{kuhn2007}
while engineering disorder with different correlation properties is possible.

The aim of this paper is to study transport in a disordered honeycomb
lattice. We first stick to the simpler case of the tight-binding model for the
regular honeycomb lattice in the presence of uncorrelated on-site disorder
characterized by a symmetric box distribution. The more interesting (but
more complicated) case of correlated on-site disorder will be the scope of a
forthcoming publication. The novelty of our work lies in the generalization
of two known methods,
the recursive Green's function method and the self-consistent Born
approximation,
(i) to extract single-particle properties, such as the scattering mean free
time $\tau$, the scattering mean free path $\ell$, the density of state
(DoS) $\nu$, and the localization length $\xi$, and (ii) to relate them
to experimentally-controlled parameters, such as the tunneling amplitude
$J$ and the disorder strength $W$. The numerical data obtained from
these two methods show remarkable agreement and give an accurate estimation
of these single-particle properties. Our results further confirm the
one-parameter scaling hypothesis\cite{abrahams1979} for localization but
also reveal a quantitative disparity with the predictions of the self-consistent theory
of localization~\cite{vollhardt1980,*wolfle2010,kuhn2007}. This disparity
is yet to be understood.

Currently, there are three pieces of numerical works that are technically
relevant to ours.
Using a transfer matrix technique, Schreiber and Ottomeier~\cite{schreiber1992}
have shown that the localization lengths for various lattices (square, honeycomb and triangular) at the energy band centre $E=0$ -- thus including the Dirac point of
the honeycomb lattice -- and for various disorder strengths obey the same scaling
laws. Using the same method, Xiong and Xiong
concluded that all states are localized but found that the scaling behavior
at the charge neutrality point is different from the one at different
energies~\cite{xiong2007}. On the other hand, Lherbier et al. considered
the time dynamics of a random phase wave packet using a real space order-$N$
Kubo method~\cite{lherbier2008}. They subsequently extracted the diffusion
constant, and hence the scattering mean free path (which coincides
with the transport mean free path as scattering by our $\delta$-correlated potential
is isotropic),
from the time evolution of the spatial spread of the wave packet. This
extracted scattering mean free path was then used to deduce the semi-classical
conductivity through Einstein-Kubo relation~\cite{kubo1957} and the
localization length through the self-consistent theory of localization,
a procedure that our numerical data question.

The paper is organized as follows. Sec.~\ref{sec:model} gives the essential
ingredients of our model and the eigenstructure of the disorder-free honeycomb
lattice in the tight-binding regime. In Sec.~\ref{sec:se}, we introduce
the self-energy and detail the Born and self-consistent Born approximations
(SCBA). We further derive analytical expressions for the self-energy at some
particular energies in the weak-disorder limit. In Sec.~\ref{sec:numeric},
we introduce the recursive Green's function (RGF) method which is designed to
compute exact matrix elements of the Green's function for large system
sizes, with the caveat that actual computations can take months on a computer
cluster. A faster but more restricted variant is the recursive transfer matrix
method. Together with a finite-size analysis, these methods allow us to extract
the localization length  of the
disordered honeycomb lattice at any given energy. In Sec.~\ref{sec:results},
we perform a finite-size analysis of the localization lengths.
We show
that they can be simultaneously scaled for all energies, including the charge
neutrality point.
Furthermore, our results indicate that this universal curve is valid for
all lattice types,
all energies within the energy band and possibly for all types of uncorrelated
disorders,
which is not surprising from the viewpoint of the one-parameter scaling
hypothesis. In addition to the localization length, we also extract
the scattering mean free path  and the DoS from the recursive Green's
function method evaluated at complex energies. The extracted quantities
show remarkable agreement with our results from the self-consistent Born
approximation. The comparison of the numerically computed localization length to the prediction
of the self-consistent theory of localization shows a fair qualitative agreement,
but marked quantitative differences. We finally conclude in Sec.~\ref{sec:conclusion}. Additional details are given
in the Appendices.

\section{Model}
\label{sec:model}
\begin{figure}
\includegraphics[width=0.45\textwidth, clip=]{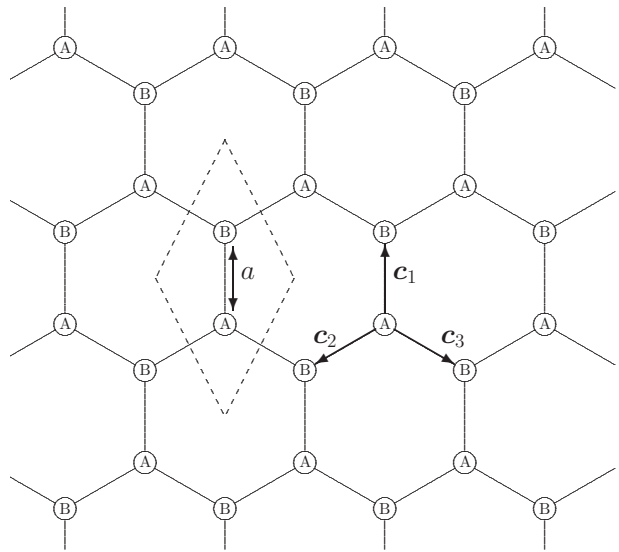}
\caption{\label{fig:latticeAll}
A honeycomb lattice with lattice constant $a$ and its diamond-shaped
two-point basis cell (dashed line). The vectors $\vec{c}_l$ ($l=1,2,3$)
connect an $\textsc{a}$-site to its three $\textsc{b}$-site nearest-neighbors.}
\end{figure}

We consider here a tight-binding Hamilton operator acting on a regular
honeycomb lattice with on-site disorder
\begin{equation}
	H = H_0 + V = -J \sum_{\langle i,j\rangle}\bigl(|i\rangle\langle
	j|+|j\rangle\langle i|\bigr) + \sum_i \varepsilon_i|i\rangle\langle
	i|,\label{eq:tightBinding}
\end{equation}
where $|i\rangle$ refers to a Wannier state localized on site $i$ and $\langle
i,j \rangle$
denotes a sum over nearest neighbors. The hopping parameter $J$ is usually
positive and it fixes the energy scale. Throughout the paper,
we assume the diagonal elements $\varepsilon_i$ to be independent random
variables characterized by the same symmetric box probability distribution
\begin{equation}
	P(\varepsilon_i)=\left\{\begin{array}{ll}
				   \frac{1}{W} & \mbox{for }|\varepsilon_i|\leq
				   \frac{W}{2},\\
				   0 & \mbox{otherwise}.
				\end{array}\right.
\end{equation}
where $W$ is the disorder strength. The disorder has thus zero mean
average $\overline{\varepsilon_i} =0$ and its two-point correlator is then
$C_{ij}=\overline{\varepsilon_i\varepsilon_j}=\frac{W^2}{12} \delta_{ij}$,
where the bar denotes averaging over disorder configurations. The disorder being spatially $\delta$-correlated, scattering is isotropic; the scattering
and transport mean free paths are consequently equal.
 
We now briefly review the eigenstructure of the disorder-free Hamiltonian
$H_0$. We refer to Ref.~\cite{lee2009} for more details. The regular honeycomb
lattice being a triangular Bravais lattice with a two-site basis cell, it can
be pictured as two shifted triangular sublattices denoted by $\textsc{a}$
and $\textsc{b}$, see Fig.~\ref{fig:latticeAll}. As a consequence, the
coordination number of the honeycomb lattice is three. We denote by $\vec{c}_l$
($l=1,2,3$) the link vectors connecting a site $i\in \textsc{a}$ to its
three nearest-neighbors $j_l \in B$ ($|\vec{c}_l| = a$, $a$ being the lattice
constant). We next define the structure factor of the honeycomb
lattice for nearest-neighbor hopping as
\begin{equation}
\label{eq:strucfac}
f(\vec{k}) = |f(\vec{k})|\ e^{\mathrm{i}\varphi(\vec{k})} =  \sum_{l=1}^{3}
e^{\mathrm{i} \vec{k}\cdot\vec{c}_l}.
\end{equation}
For the honeycomb lattice depicted in Fig.~\ref{fig:latticeAll}, where $\vec{c}_1$ points along the $y$-axis, we have
\begin{equation}
|f(\vec{k})|^2 =
1+4\cos^2\left(\frac{\sqrt{3}k_xa}{2}\right)+4\cos\left(\frac{\sqrt{3}k_xa}{2}\right)\cos\left(\frac{3k_ya}{2}\right).
\end{equation}

\begin{figure}
\includegraphics[width=0.45\textwidth]{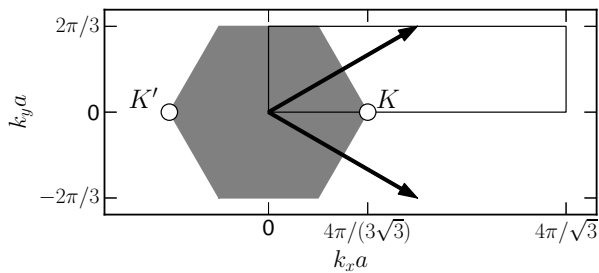}
\caption{\label{fig:bz}
The first Brillouin zone $\mathcal{B}$ (in grey) of the honeycomb lattice shown in Fig.~\ref{fig:latticeAll} and the
associated reciprocal lattice vectors (arrows). The two empty circles mark
the Dirac points, around which the energy dispersion is linear. By using
convenient reciprocal lattice translations, the first Brillouin zone can be
mapped onto the shown rectangle which is then used as the integration space
in all integrals in the paper.}
\end{figure}

As shown in Ref.~\cite{lee2009}, the eigenstructure of $H_0$ is defined by
\begin{align}
& H_0=\sum_{\vec{k}\in\mathcal{B},s=\pm} \, \varepsilon_{\vec{k}s} \,
|\vec{k}s\rangle\langle\vec{k}s|, \\
& \varepsilon_{\vec{k}s} = s \, \varepsilon_{\vec{k}} = s \, J\, |f(\vec{k})|,
\\
& |\vec{k}s\rangle = \frac{1}{\sqrt{2N_c}} \, \big[ \sum_{i\in A} e^{\mathrm{i}
\vec{k}\cdot\vec{r}_i} |i\rangle - s \, e^{-\mathrm{i} \varphi(\vec{k})}
\, \sum_{j\in B} e^{\mathrm{i} \vec{k}\cdot\vec{r}_j} |j\rangle \big]
\label{eq:ks}
\end{align}
where $\mathcal{B}$ is the first Brillouin zone of the honeycomb lattice
(see Fig.~\ref{fig:bz}), $s$ is the band index ($s=+1$ for the upper, or
conduction, band; $s=-1$ for the lower, or valence, band) and $N_c$ is the
number of Bravais cells of the lattice.

The full spectrum span the energy interval $[-3J,3J]$ and band crossing can
only occur at zero energy since $\varepsilon_+(\vec{k})=\varepsilon_-(\vec{k})$
implies that $f(\vec{k})=0$. This happens at the Dirac points
$\vec{K}$ and $\vec{K}'=-\vec{K}$ where
\begin{equation}
 \vec{K} = \left( \frac{4\pi}{3\sqrt{3}a}, 0 \right)
\label{eq:Dirac}
\end{equation}
for the honeycomb lattice in Fig.~\ref{fig:latticeAll}. In the solid-state community, the energy at
the Dirac points is usually referred to as the charge neutrality (energy) point
because the number of energy states above and below this point are equal. As
a consequence, when the gate voltage is fixed at $\varepsilon_\pm(\vec{K})=0$
in a graphene sample, the graphene sheet is charge neutral since the particle
and hole states are balanced.

In the rest of the paper, we will be mainly interested in the different
localization properties of our model for energies $E$ lying near the band
edges, $E\approx \pm 3J$, and near the band centre, $E\approx 0$. In the
first case, $\vec{k}$ lies near the centre of the Brillouin zone ($ka \ll
1$, where $k=|\vec{k}|$), while in the second case $\vec{k}$ lies near the
Dirac points ($qa \ll 1$, where $\vec{q}= \vec{k}-\vec{K}$, and similarly
around $\vec{K}'$).

Near the band edges $E =\pm 3J$, the dispersion relation is quadratic,
thus representing free massive particles
\begin{equation}
 \varepsilon_{\vec{k}s}\approx 3J s \left(1-\frac{a^2k^2}{4} \right)\ \ \
 \mbox{$ka\ll1$},
\label{eq:enDisperse1}
\end{equation}
the effective mass $m_s = -s \, 2 \hbar^2/(3Ja^2)$ being negative for the
upper band and positive for the lower band.

Near the charge neutrality point $E=0$, the dispersion relation is linear,
thus representing the celebrated ``relativistic" massless Dirac particles
propagating with a velocity $c$ playing the role of an effective speed
of light,
 \begin{eqnarray}
 &\varepsilon_{\vec{k}s}& \, \approx s	\frac{3J}{2}aq = s\hbar c q\ \ \ \
 \mbox{ ($qa \ll 1$)},\\
 &c& = \frac{3aJ}{2\hbar}.
 \label{eq:enDisperse2}
\end{eqnarray}

\section{Self-energy}
\label{sec:se}

An initial Bloch state $|\vec{k}s\rangle$ propagating in the lattice will
suffer scattering by the disorder fluctuations and thus will be depleted,
decaying exponentially over time with a time constant which is the scattering
mean free time $\tau_{\vec{k}s}$. This coherent propagation and decay are
described by the disorder-averaged Green's function $\overline{G}$ which
obey the Dyson's equation
 \begin{equation}
	\overline{G}(z)=G_0(z) + G_0(z)\Sigma(z)\overline{G}(z),
\end{equation}
$z$ being a point in the complex energy plane. Within our model, the
disorder-free Green's function is given by
\begin{equation}
\label{eq:G0}
G_0(z) = \frac{1}{z-H_0} = \sum_{\vec{k}s} \
\frac{|\vec{k}s\rangle\langle\vec{k}s|}{z-\varepsilon_{\vec{k}s}}.
\end{equation}
The Dyson equation features the self-energy $\Sigma$ which
is a central quantity given by a perturbative sum of irreducible
diagrams~\cite{akkermans2007}. As the disorder average restores the lattice
translation symmetries and characteristics, one has
\begin{eqnarray}
\langle\vec{k}'s'|\overline{G}(z)|\vec{k}s\rangle = \overline{G}_{\vec{k}s}(z)
\ \delta_{\vec{k}\vec{k}'} \ \delta_{ss'}, \\
\langle\vec{k}'s'|\Sigma(z)|\vec{k}s\rangle = \Sigma_{\vec{k}s}(z) \
\delta_{\vec{k}\vec{k}'} \ \delta_{ss'},
\end{eqnarray}
where the dependence of  $\Sigma_{\vec{k}s}(z)$ on $\vec{k}$ is usually
smooth. Due to particle-hole symmetry, the spectrum of $H_0$ is symmetric
with respect to $E=0$. Since the on-site energies are themselves independent
symmetrically-distributed random variables, it is easy to show that $\langle \vec{k}s |\overline{G}(z)|  \vec{k}s \rangle^*=
-\langle \vec{k},-s | \overline{G}(-z^*) | \vec{k},-s \rangle$, where the
star stands for complex conjugation.
In turn, the self-energy satisfies 
\footnote{These identities no
longer hold for a speckle potential since it breaks the $V \to -V$ symmetry.}

\begin{align}
\Sigma_{\vec{k},s}(z) = \Sigma^*_{\vec{k},-s}(-z^*).\label{eq:propsig}
\end{align}
Furthermore,
because of time-reversal invariance, the self-energy is the same at $\pm
\vec{k}$. This means that it is sufficient to study the negative energy sector
and forward propagation ($k_x \geq 0$). 
The scattering mean free time, defined through
\begin{equation}
\frac{\hbar}{2\tau_{\vec{k}s}(E)} = - \ \textrm{Im}\Sigma_{\vec{k}s}(E),
\end{equation}
is independent of the band index.

\subsection{Born approximation}

An analytical expression for $\Sigma_{\vec{k}s}(z)$ is generally not available
and one has to resort to approximations to find the self-energy. For weak
disorder, the simplest approximation is the Born approximation which consists
in discarding all terms of the full diagrammatic perturbative expansion
in Fig.~\ref{fig:selfE_BA}
except the first one. Its lattice matrix elements are
\begin{equation}
\label{eq:ba1}
\langle i| \Sigma_{\textrm{Born}}(z)|j\rangle = C_{ij} \  \langle i| G_0(z)
|j\rangle = \frac{W^2}{12} \ I(z) \ \delta_{ij}
\end{equation}
with
\begin{equation}
I(z) =	\langle i| G_0(z) |i\rangle = \int_{\mathcal{B}}
\frac{d\vec{k}}{\Omega} \ \frac{z}{z^2-J^2|f(\vec{k})|^2},
\end{equation}
where $\Omega = 8\pi^2/(3\sqrt{3}a^2)$ is the area of the first Brillouin
zone. For uncorrelated on-site disorder, the self-energy
at the Born approximation is a scalar:
\begin{equation}
\Sigma_{\vec{k}s}(z) \approx \Sigma_{\textrm{Born}}(z) = \frac{W^2}{12} \ I(z).
\end{equation}
The average Green's function then reads
\begin{equation}
\overline{G}_{\textrm{Born}}(z) = \frac{1}{z-H_0-\Sigma_{\textrm{Born}}(z)}
= G_0(z-\Sigma_{\textrm{Born}}(z)).
\end{equation}
The expression of $I(z)$ in terms of elliptic integrals~\cite{morita1971,horiguchi1972} is given in
Appendix~\ref{AppendixG0}.

\subsection{Self-consistent Born approximation (SCBA)}
\label{sec:scba}
\begin{figure}
\includegraphics[width=0.49\textwidth]{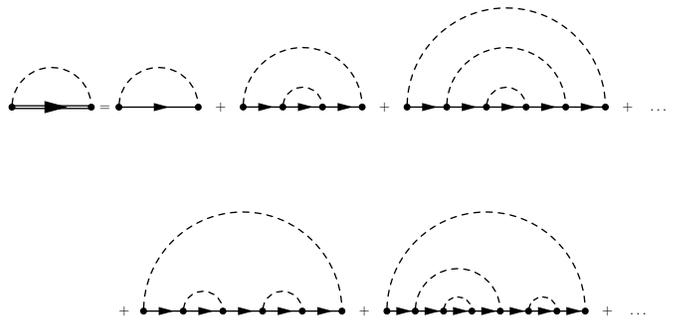}
\caption{\label{fig:selfE_BA}
The ``rainbow" subclass of diagrams retained to compute the self-energy in
the self-consistent Born approximation. The double line with arrow denotes
an averaged Green's function $\overline{G}$ while a single line with arrow
denotes a disorder-free lattice Green's function $G_0$.
Two vertices (solid dots) connected by a dashed line represent a 2-point
correlator $C_{ij} = \overline{\varepsilon_i\varepsilon_j}$. The Born
approximation consists in computing the self-energy with the first ``rainbow"
diagram only. For uncorrelated disorder, the connected vertices correspond
thus to the same site. In this case the self-energy is a scalar operator depending only on the
energy and the disorder strength.}
\end{figure}

This approximation scheme builds on the Born approximation by replacing $G_0$
by $\overline{G}$ in Eq.~\eqref{eq:ba1}. It is more powerful as it amounts to sum the infinite subclass
of ``rainbow" diagrams given in Fig.~\ref{fig:selfE_BA}. It gives the following self-consistent equation:
\begin{equation}
\Sigma_{\textrm{SCBA}}(z) = 
\frac{W^2}{12} \ \langle i| \overline{G}(z) |i\rangle = \frac{W^2}{12} \ I(z-\Sigma_{\textrm{SCBA}}(z)),
\label{eq:scba1}
\end{equation}
which is easy to solve numerically.

\begin{figure}
\includegraphics[width=0.49\textwidth]{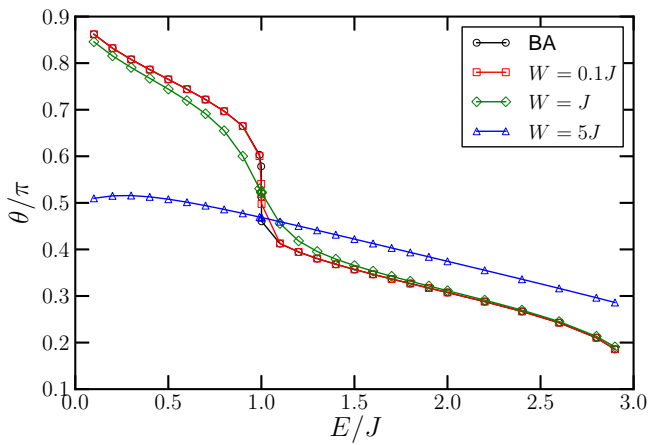}
\caption{\label{fig:SCBA_theta} (Color online) Using the parametrization $\Sigma(z) = \gamma J \Exp{-\mri \theta}$ ($\gamma
\geq 0$, $0\leq \theta \leq \pi$), the figure shows the angle $\theta$ as
a function of the energy $E$ in the Born approximation (black open circles)
and in the SCBA for various disorder strengths. In the Born approximation,
$\theta$ is independent of $W$. At $W=0.1J$, SCBA and the Born approximation are
essentially identical.}
\end{figure}

\begin{figure}
\includegraphics[width=0.49\textwidth]{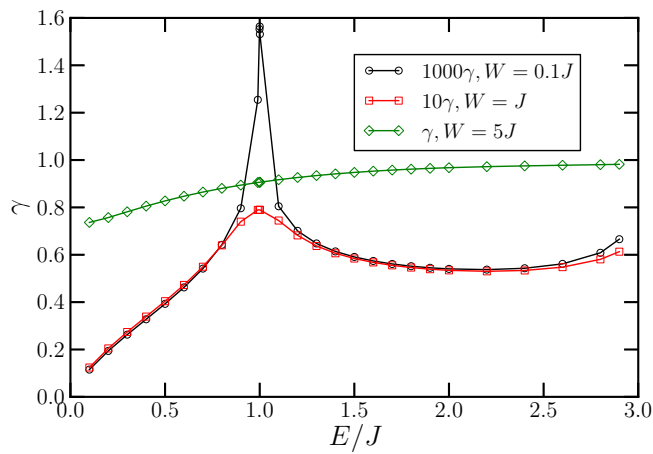}
\caption{\label{fig:SCBA_gamma} (Color online) Using the parametrization $\Sigma(z) = \gamma J \Exp{-\mri \theta}$ ($\gamma
\geq 0$, $0\leq \theta \leq \pi$), the figure shows the amplitude $\gamma$ as a
function of the energy $E$ in the SCBA for various disorder strengths. As $W\to 0$, the peak at $E=J$ develops
into the van Hove singularity.}
\end{figure}

\begin{figure}
\includegraphics[width=0.49\textwidth]{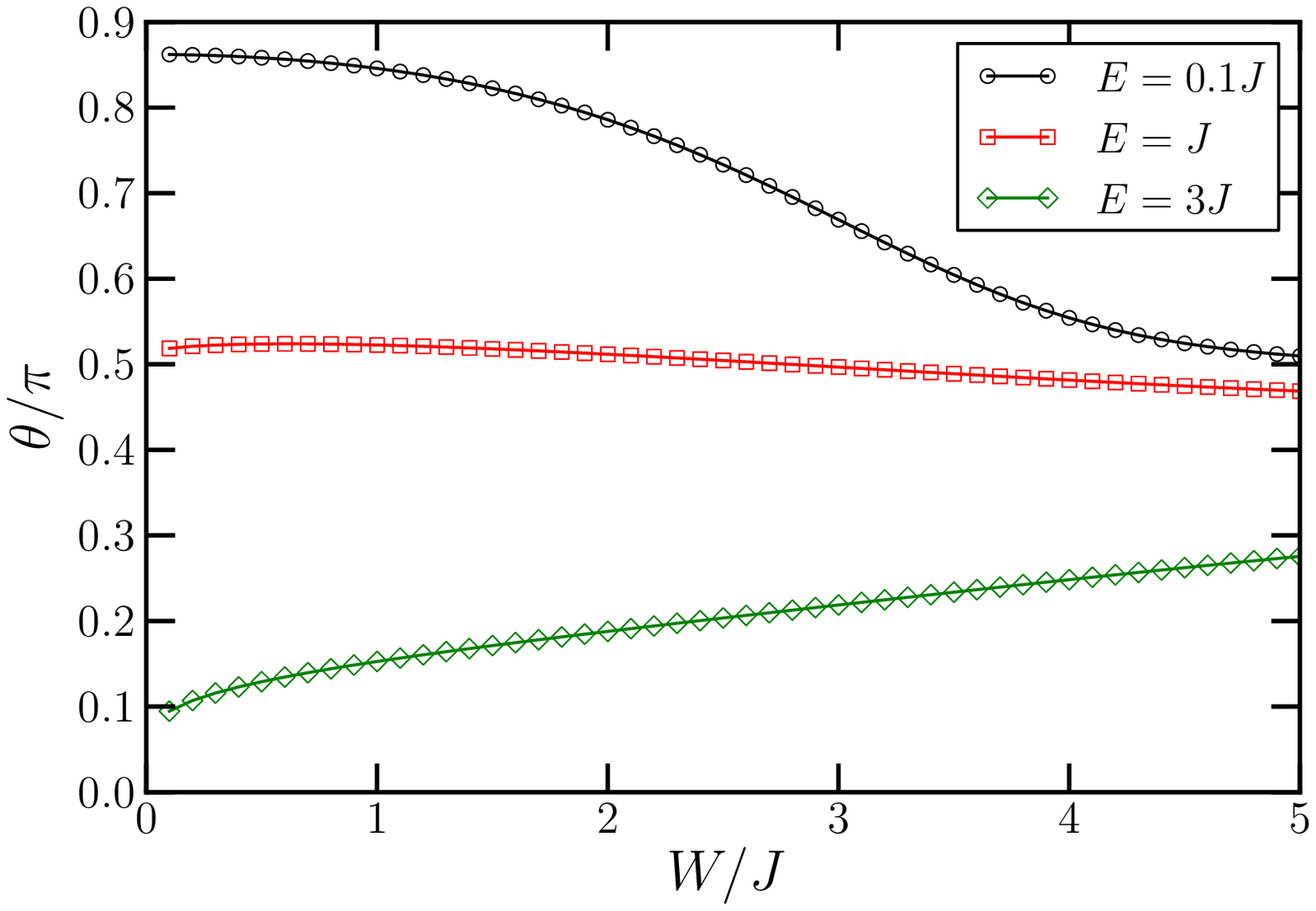}
\caption{\label{fig:SCBA_theta_W}
(Color online) Using the parametrization $\Sigma(z) = \gamma J \Exp{-\mri \theta}$ ($\gamma
\geq 0$, $0\leq \theta \leq \pi$), the figure shows the angle $\theta$ as a
function of the disorder strength $W$ in the SCBA near the charge neutrality
point, at the van Hove singularity and at the band edge.}
\end{figure}

\begin{figure}
\includegraphics[width=0.49\textwidth]{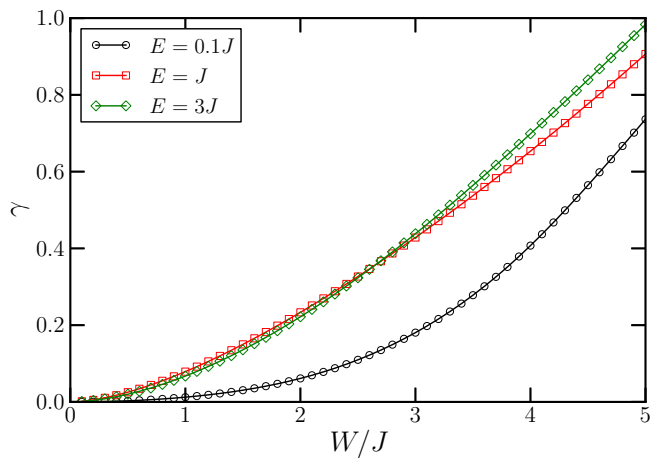}
\caption{\label{fig:SCBA_gamma_W}
(Color online) Using the parametrization $\Sigma(z) = \gamma J \Exp{-\mri \theta}$ ($\gamma
\geq 0$, $0\leq \theta \leq \pi$), the figure shows the amplitude $\gamma$
as a function of the disorder strength $W$ in the SCBA near the charge
neutrality point, at the van Hove singularity and at the band edge.}
\end{figure}

In the following we will use the parametrization $\Sigma_{\textrm{SCBA}}(z)
= \gamma J e^{-\mri \theta}$ with $\gamma$ positive. For the scattering time
to be positive, we must have  ${\rm Im}\Sigma_{\vec{k}s} <0$, which enforces
$0\leq \theta \leq \pi$. Eq.~\eqref{eq:propsig} then translate into
\begin{eqnarray}
\label{eq:propgamtheta}
\gamma(-E,W) &=& \gamma(E,W) \label{eq:propgam} \\
\theta(-E,W) &=& \pi - \theta(E,W). \label{eq:proptheta}
\end{eqnarray}
This implies $\theta(0,W) =\pi/2$ for any disorder strength
$W$. Figs.~\ref{fig:SCBA_theta} and \ref{fig:SCBA_gamma} show
the $E$-dependence of $\theta$ and $\gamma$ in the SCBA for some
particular disorder strengths $W$, while Figs.~\ref{fig:SCBA_theta_W} and
\ref{fig:SCBA_gamma_W} show the $W$-dependence of $\theta$ and $\gamma$
in the SCBA at some particular energies $E$. In the following subsection,
we investigate SCBA analytically in the weak disorder regime.

\subsection{Weak disorder limit}

In the weak disorder regime $W\ll J$, one expects $\gamma \ll 1$. Several
analytical results can then be derived in this limit. Some details are
exposed in Appendix~\ref{AppendixSCBA}.

\subsubsection{Charge neutrality point}
At the charge neutrality point, we have $z=E=0$ and the function
$I$ in~\eqref{eq:scba1} is thus evaluated at the dimensionless
complex number $Z=- \Sigma_{\textrm{SCBA}}(0,W)/J$. Furthermore,
because of Eq.~\eqref{eq:propsig}, $\theta(0,W) = \pi/2$, and
$\Sigma_{\textrm{SCBA}}(0,W) = -\mri \gamma(0,W) J$ is purely
imaginary. Equations~\eqref{eq:scba1} and \eqref{eq:Izfull} have two
solutions. One is the trivial solution $\gamma(0,W)=0$ while the nontrivial
solution solves
\begin{equation}
\int_{\mathcal{B}} \frac{d\vec{k}}{\Omega}
\frac{1}{|f(\vec{k})|^2+\gamma^2(0,W)}= \frac{12J^2}{W^2}.
\end{equation}
An expansion of the elliptic integrals appearing in the first line of
Eq. \eqref{eq:Izfull} for $|Z| \ll 1$ leads to
\begin{equation}
\label{eq:scba0}
	\gamma(0,W) \approx 3 \exp\big(-\frac{6\pi\sqrt{3}J^2}{W^2}\big).
\end{equation}
Shon {\it et al.} found essentially the same type of result $\gamma
J=\varepsilon_c\exp(-A_1/W^2)$, see Eq.~(3.21) in~\cite{shon1998}. The
difference between their parameter values and ours arises from their introduction
of the cut-off energy $\varepsilon_c$. In our treatment, we use the exact dispersion relation, beyond the linear approximation around the Dirac points,
making the introduction of an artificial cut-off unnecessary.

\subsubsection{Van Hove singularities}
At the van Hove singularities, the disorder-free DoS diverges. For our honeycomb
tight-binding model, this occurs at $z=E=\pm J$~\cite{lee2009} and the function
$I$ in~\eqref{eq:scba1} is now evaluated at the dimensionless complex number
$Z_+= 1 -\Sigma_{\textrm{SCBA}}(J,W)/J$. A small-parameter expansion of
equations~\eqref{eq:scba1} and \eqref{eq:Izfull} around $Z = 1$ now leads to
\begin{align}
	\gamma(J,W) &\approx&  \frac{W^2}{16\pi J}\left(\ln\Bigl(\frac{64\pi
	J^2}{W^2}\Bigr)+\ln\ln\Bigl(\frac{64\pi J^2}{W^2}\Bigr)\right),
	\label{eq:vHgam} \\
	\theta(J,W) &\approx& \frac{\pi}{2} \left( 1+
	\frac{1}{3\big(\ln(\frac{64\pi J^2}{W^2}) - \ln\ln(\frac{64\pi
	J^2}{W^2})\big)} \right).  \label{eq:vHtheta}
\end{align}

One sees that, at lowest order, the self-energy at the van Hove singularities
is purely imaginary too, $\theta\approx\pi/2$.

\subsubsection{Band edges}
At the disorder-free band edges, $z=E=\pm 3J$. By the same token, we evaluate the function
$I$ in~\eqref{eq:scba1} at the dimensionless complex number $Z_+=3 -
\Sigma_{\textrm{SCBA}}(3J,W)/J$. A small-parameter expansion around $Z= 3$
now leads to
\begin{align}
	\Sigma_{{\rm SCBA}}(3J,W) \approx  &\frac{\sqrt{3}W^2}{48\pi J} \Big(
	\ln\Bigl( \frac{192\sqrt{3}\pi J^2}{W^2}\Bigr) - \mri\pi \nonumber\\
	&-\ln(\ln\Bigl( \frac{192\sqrt{3}\pi J^2}{W^2}\Bigr) - \mri\pi)\Big).
\end{align}
Anticipating results displayed and discussed in Paragraph \ref{SubSec:DoS}, within the SCBA scheme, the DoS vanishes outside a finite energy band, with a square-root behavior near the band edge. This SCBA band edge is approximately given by the solution of the equation $E - \textrm{Re}\Sigma_{\textrm{SCBA}}(E) =3J$ and does not coincide with the exact band edge $3J+W/2$ found for the box disorder. Note also that the Lifshitz tail between $3J$ and $3J+W/2$ is completely missed by the SCBA scheme.

\section{Recursive Green's function method}\label{sec:numeric}

The RGF method is based upon the division of the system in smaller
sections for which the Green's functions can be calculated more
easily~\cite{mkk1983}. These sections are then ``glued together" one
after one and Dyson's equation~\cite{mahan2000} is repeatedly used to
derive the full Green's function in terms of the Green's functions of
the smaller sections. To study localization, we will use a generalized
version~\cite{mackinnon1985,pohlmann1988} of the RGF method. This generalized
version enables us to extract {\it any} lattice matrix element of the Green's
function conveniently and with high numerical stability.

\begin{figure}[!ht]
	\includegraphics[width=0.4\textwidth]{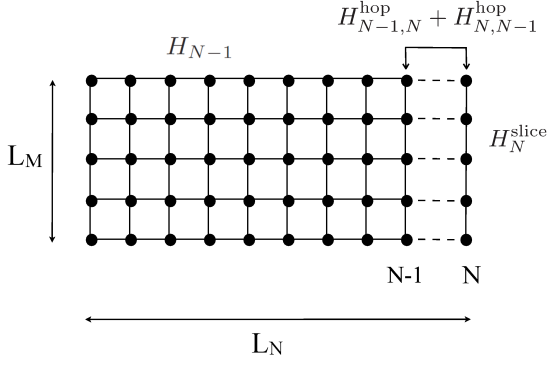}
		\caption{
\label{fig:recG}
Schematic illustration of the RGF method with a strip of square lattice of
length $L_N$ and width $L_M$. The system under consideration is constructed by
repeated stacking of vertical slices of length $L_M$. The Green's function at
each step of the stacking process is calculated recursively using Dyson's
equation. Knowing the Green's function $G^{(N-1)}=(E+\mri 0^+-H_{N-1})^{-1}$ 
for the system with
$(N-1)$ slices, the hop operator $H_N^{\rm hop}= H^{\rm  hop}_{N-1,N}+H^{\rm
hop}_{N,N-1}$ coupling slices $(N-1)$ and $N$, and the Hamilton operator
$H_N^{{\rm slice}}$ of the isolated slice $N$, one can exactly compute
the Green's function $G^{(N)}$ for the whole system of $N$ slices, see
Eqs.~\eqref{eq:rgf}. }
\end{figure}

Applying the RGF scheme to our case amounts to consider a finite quasi-1D
lattice strip and to divide it into $N$ vertical slices, the two open ends
being along the horizontal direction, see Fig.~\ref{fig:recG}. Denoting
by $H_{N-1}$ the nearest-neighbor tight-binding Hamilton operator for a
strip with $(N-1)$ slices, the Hamilton operator $H_N$ obtained by gluing
an additional slice $N$ can be split into three terms,
\begin{equation}
	H_N = H_{N-1} + H_N^{\rm  slice} + H^{\rm  hop}_{N-1,N} + H^{\rm
	hop}_{N,N-1},
	\label{eq:splitH_N}
\end{equation}
where $H^{\rm  hop}_{N-1,N}$ is the nearest-neighbor hop operator connecting
sites within slice $(N-1)$ to sites within slice $N$ (and vice-versa for
$H^{\rm  hop}_{N,N-1}$). Since no external gauge fields are present in our
model, we safely consider the hop operators to be real from now on. $H_N^{\rm
slice}$ is the nearest-neighbor tight-binding Hamilton operator for the
isolated slice $N$ before it is stacked to the others. It thus includes the
on-site disorder diagonal term $V$
in Eq.~\eqref{eq:tightBinding}.

Using Dyson's equation, the submatrix $G^{(N)}_{l,n}$
of the full retarded Green's  function $G^{(N)}=(E+\mri 0^+-H_N)^{-1}$
coupling slice $l$ to slice $n$ at energy $E$ can be obtained through the
following recursion relations,
\begin{widetext}
\begin{subequations}
\label{eq:rgf}
\begin{align}
G^{(N)}_{l,n} &=G^{(N-1)}_{l,n} + G^{(N-1)}_{l,N-1} \ H^{\rm hop}_{N-1,N} \ G^{(N)}_{N,n},\label{eq:recG1}\\
G^{(N)}_{l,N} &= G^{(N-1)}_{l,N-1} \ H^{\rm  hop}_{N-1,N} \ G^{(N)}_{N,N},\label{eq:recG2}\\
G^{(N)}_{N,n} &= G^{(N)}_{N,N} \ H^{\rm  hop}_{N,N-1} \ G^{(N-1)}_{N-1,n},\label{eq:recG3}\\
G^{(N)}_{N,N} &= \biggl(E + \mri 0^+ - H_N^{\rm  slice} - H^{\rm	hop}_{N,N-1} \ G^{(N-1)}_{N-1,N-1} \ H^{\rm  hop}_{N-1,N} \biggr)^{-1}\label{eq:recG4}
\end{align}
\end{subequations}
\end{widetext}
with $1\leq l \leq n \leq (N-1)$, see Appendix \ref{sec:recG_derive} for a
short derivation.

In the following subsections, we explain how we chose to slice
the honeycomb lattice for the zigzag and armchair geometries. Once the
stacking is properly defined, each lattice site $i \equiv (n,m)$ will then
be labelled by two integers. The first one $1\leq n \leq N$ corresponds to
the slice it belongs to from left to right, $N$ being the total number of
such slices. The second one $1 \leq m \leq M$ corresponds to its position
along the slice from bottom to top, $M$ being the total number of sites per
slice. For both geometries we have checked that the matrix elements of the
Green's function obtained through the recursive algorithm agree well with
those computed by direct inversion of the operator $(E+\mri 0^+ -H_N)$. 

\subsection{Zigzag configuration}
\label{ZZ}

\begin{figure}[!ht]\centering
	\includegraphics[width=0.4\textwidth]{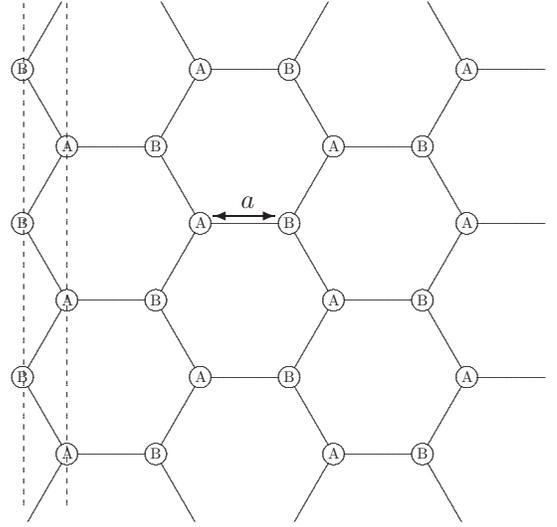}
\caption{
\label{fig:latticeZZ}
In the honeycomb zigzag configuration with periodic boundary conditions along
the vertical direction, $L$ zigzag vertical chains, each containing $2M$
sites, are stacked along the horizontal direction ($L=4$ and $M=3$ in the
figure). Each zigzag chain defines two vertical slices (dashed vertical lines).
These slices are labeled from left to right by the integer $1 \leq n \leq
N=2L$. Each slice contains $M$ sites labeled from bottom to top by the integer
$1 \leq m \leq M$. The slice gender alternates between \textsc{a}-type and
\textsc{b}-type. Each site in the lattice is uniquely parametrized by the
couple of integers $(n,m)$.}
\end{figure}

We first consider the zigzag (ZZ) configuration depicted in
Fig.~\ref{fig:latticeZZ}, where $L$ vertical zigzag chains of the honeycomb
lattice are stacked along the horizontal direction. In the following we
will be essentially interested in the case of periodic boundary conditions
in the vertical direction and open boundary conditions in the horizontal
one. This imposes the number of sites in each zigzag chain to be an
even integer $2M$. For each zigzag chain, one can define \textsc{a}-type
(resp.	\textsc{b}-type) vertical slices containing only \textsc{a}-sites
(resp. \textsc{b}-sites). Each of these vertical slices contain $M$ sites
and there are $N=2L$ such slices,  \textsc{a}-type slices alternating with
\textsc{b}-type slices. Lattice sites are thus parametrized by $(n,m)$ with
$1\leq n \leq N=2L$ and $1\leq m \leq M$, see Fig.~\ref{fig:latticeZZ}. The
width and length of this honeycomb ZZ strip are $L_M=\sqrt{3}Ma$ and $L_N
\approx 3Na/4$ ($N\gg 1$).

With this slicing choice, the Hamiltonian $H_n^{\rm  slice}$ associated to the
isolated slice $n$ is simply a $M\times M$ diagonal matrix with entries given
by the $M$ on-site disorder elements $\{\varepsilon_{n,m}\}$. Furthermore,
one can easily see that
\begin{align}
&H_{n,n+1}^{\rm  hop} =-J \, \mathbb{1}  \hspace{0.5cm}  &{\rm if} \, n \equiv 0 \, [{\rm mod.} 4] \nonumber \\
&H_{n,n+1}^{\rm  hop} =-J \, {\bf D}  \hspace{0.5cm}  &{\rm if} \, n \equiv 1 \, [{\rm mod.} 4] \nonumber \\
&H_{n,n+1}^{\rm  hop} =-J \, \mathbb{1}  \hspace{0.5cm}  &{\rm if} \, n \equiv 2 \, [{\rm mod.} 4] \nonumber \\
&H_{n,n+1}^{\rm  hop} =-J \, {\bf D}^{\rm T}  \hspace{0.5cm}  &{\rm if} \, n \equiv 3 \, [{\rm mod.} 4] \nonumber \\
&H_{n+1,n}^{\rm  hop} = (H_{n,n+1}^{\rm  hop})^{\rm T} \nonumber \\
\end{align}
where the ${\rm T}$-superscript means matrix transposition and ${\bf D}$
is the $M\times M$ matrix given by
\begin{equation}
	{\bf D} = \begin{pmatrix}
		  1 & 0 & 0 & \cdots & 0 & p\\
		  1 & 1 & 0 & \cdots & 0 & 0\\
		  0 & 1 & 1 & \cdots & 0 & 0\\
		  \vdots & \vdots & \vdots & \ddots & \vdots & \vdots\\
		  0 & 0 & 0 & \cdots & 1 & 1
		  \end{pmatrix}.
\end{equation}
When periodic boundary conditions in the vertical direction are used (which
is our case), $p=1$. For open boundary conditions in the vertical direction,
one would have $p=0$.

An astute reader may have noticed that the determinant of ${\bf D}$ vanishes
for $M$ even and periodic boundary conditions. In this case the transfer
matrix method~\cite{mkk1983} cannot be implemented as it would require the
inversion of ${\bf D}$ or ${\bf D}^{\rm T}$. To avoid this pitfall, one
can nevertheless always choose $M$ odd. Note however that the RGF scheme is
perfectly immune to this breakdown and its implementation does not suffer
any flaw as we have duly checked.

\subsection{Armchair configuration}
\label{AC}

\begin{figure}[!ht]\centering
	\includegraphics[width=0.45\textwidth]{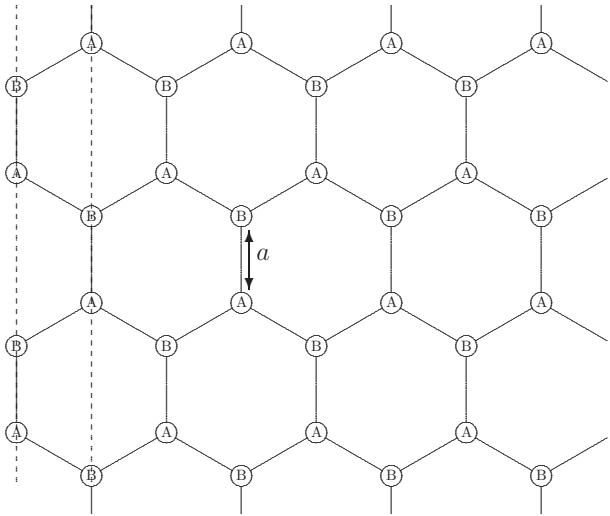}
\caption{\label{fig:latticeAC}
In the armchair configuration (AC), $M$ horizontal zigzag chains, each
containing $N$ sites, are stacked vertically ($M=4$ and $N=8$ in the
figure). There are thus $N$ vertical slices labeled from left to right by the
integer $1 \leq n \leq N$ (dashed lines). Each vertical slice contains $M$
sites labeled from bottom to top by the integer $1 \leq m \leq M$. If periodic
boundary conditions are imposed along the vertical direction, $M$ must be
even. The site gender within a slice alternates between \textsc{a}-type and
\textsc{b}-type. Each site in the lattice is uniquely parametrized by the
couple of integers $(n,m)$.}
\end{figure}

We now turn to the honeycomb armchair configuration where $M$ horizontal
zigzag chains, each containing $N$ sites, are stacked along the vertical
direction, see Fig.~\ref{fig:latticeAC}. When periodic boundary conditions
are imposed along the vertical direction, $M$ must be even. The left vertical
boundary of the lattice is now reminiscent of the shape of an armchair. Using
the same recipe, we slice the lattice with vertical lines. There are now $N$
such vertical slices, each containing $M$ sites. The width and length of this
honeycomb AC strip are $L_M=3Ma/2$ and $L_N \approx \sqrt{3}Na/2$ ($N\gg 1$).

In the armchair configuration, it is easy to see that the $M\times M$
hop matrices satisfy $H_{n, n+1}^{\rm  hop}= H_{n+1,n}^{\rm  hop} = -J \,
\mathbb{1}$, while $H_n^{\rm slice} = - J \, {\bf X}_n + \bs{\varepsilon}_n$
where $\bs{\varepsilon}_n$ is a $M\times M$ diagonal matrix with entries equal
to the $M$ on-site disorder $\{\varepsilon_{n,m}\}$ in slice $n$. ${\bf X}_n$
is a $M\times M$ sparse matrix that couples each site to its nearest neighbors
within slice $n$, namely
\begin{equation}
{\bf X}_n=\left\{\begin{array}{ll}
\begin{pmatrix}
0\;\; & 1\;\; & 0\;\; & 0\;\; & \cdots & 0\;\; & 0\\
1\;\; & 0\;\; & 0\;\; & 0\;\; & \cdots & 0\;\; & 0\\
0\;\; & 0\;\; & 0\;\; & 1\;\; & \cdots & 0\;\; & 0\\
0\;\; & 0\;\; & 1\;\; & 0\;\; & \cdots & 0\;\; & 0\\
\vdots\;\; & \vdots\;\; & \vdots\;\; & \vdots\;\; & \ddots & \vdots\;\; & \vdots\\
0\;\; & 0\;\; & 0\;\; & 0\;\; & \cdots & 0\;\; & 1\\
0\;\; & 0\;\; & 0\;\; & 0\;\; & \cdots & 1\;\; & 0\\
\end{pmatrix}  & \mbox{if $n$ is odd,}\\
\begin{pmatrix}
0\;\; & 0\;\; & 0\;\; & 0\;\; & 0\;\; & \cdots & 0\;\; & 0\;\; & p\\
0\;\; & 0\;\; & 1\;\; & 0\;\; & 0\;\; & \cdots & 0\;\; & 0\;\; & 0\\
0\;\; & 1\;\; & 0\;\; & 0\;\; & 0\;\; & \cdots & 0\;\; & 0\;\; & 0\\
0\;\; & 0\;\; & 0\;\; & 0\;\; & 1\;\; & \cdots & 0\;\; & 0\;\; & 0\\
0\;\; & 0\;\; & 0\;\; & 1\;\; & 0\;\; & \cdots & 0\;\; & 0\;\; & 0\\
\vdots\;\; & \vdots\;\; & \vdots\;\; & \vdots\;\; & \vdots\;\; & \ddots & \vdots\;\; & \vdots\;\; & \vdots\\
0\;\; & 0\;\; & 0\;\; & 0\;\; & 0\;\; & \cdots & 0\;\; & 1\;\; & 0\\
0\;\; & 0\;\; & 0\;\; & 0\;\; & 0\;\; & \cdots & 1\;\; & 0\;\; & 0\\
p\;\; & 0\;\; & 0\;\; & 0\;\; & 0\;\; & \cdots & 0\;\; & 0\;\; & 0\\
\end{pmatrix} & \mbox{if $n$ is even,}
\end{array}\right.
\end{equation}
where $p=1$ for periodic boundary conditions along the vertical direction
(and $M$ even) and $p=0$ for open ones.

\section{Numerical results}\label{sec:results}

\subsection{Localization length $\xi$}
\label{subsec:xi}
Localized wave functions are expected to decrease exponentially at large
distances. We thus compute the localization length along the horizontal
direction as \footnote{Some authors define the localization length
$\lambda_M$ through $\overline{\ln|\psi|}$ (as we do here) or through
$\overline{\ln|\psi|^2}$. There is a factor of $2$ between these two possible
definitions.}
\begin{equation}
\frac{2}{\lambda_M}=-\lim_{N\to\infty}\frac{1}{L_N} \overline{\ln \tr \left(J^2
\ G^{(N)}_{1,N} [G^{(N)}_{1,N}]^\dag \right)},
\label{eq:loclen}
\end{equation}
where the bar indicates the average over disorder configurations. Note that it
corresponds to the log-averaged transmission in quasi-1D systems which has the
nice property of being additive and self-averaging when new slices are added~\cite{muller2011}.

This finite-size localization length $\lambda_M$ depends on the energy $E$ and
the disorder strength $W$, but also on the lattice configuration (ZZ or AC) and
width $L_M$. In all our simulations, we have used a sufficiently large number
of randomly-generated disorder configurations for each set of parameters ($E$,
$W$, $M$ and lattice configuration) such that the estimated relative error
in computing $1/\lambda_M$ is less than $0.2\%$. Furthermore $1/\lambda_M$
is always computed with lengths $L_N$ greater than $5\lambda_M$. This means
that larger number of samples are required as $\lambda_M$ increases.
To avoid numerical underflow, a rescaling of $G^{(N)}_{1,N}$ is done through
Eq.~\eqref{eq:recG2} every 10 multiplications approximately. RGF equations
require one matrix inversion per slice. To speed up the computation for the
AC lattice we switched to the recursive transfer matrix method~\cite{mkk1983}
where a true matrix inversion is carried out only after the transfer matrix
equation is applied for 10 slices.

\begin{figure*}
	\includegraphics[width=0.99\textwidth]{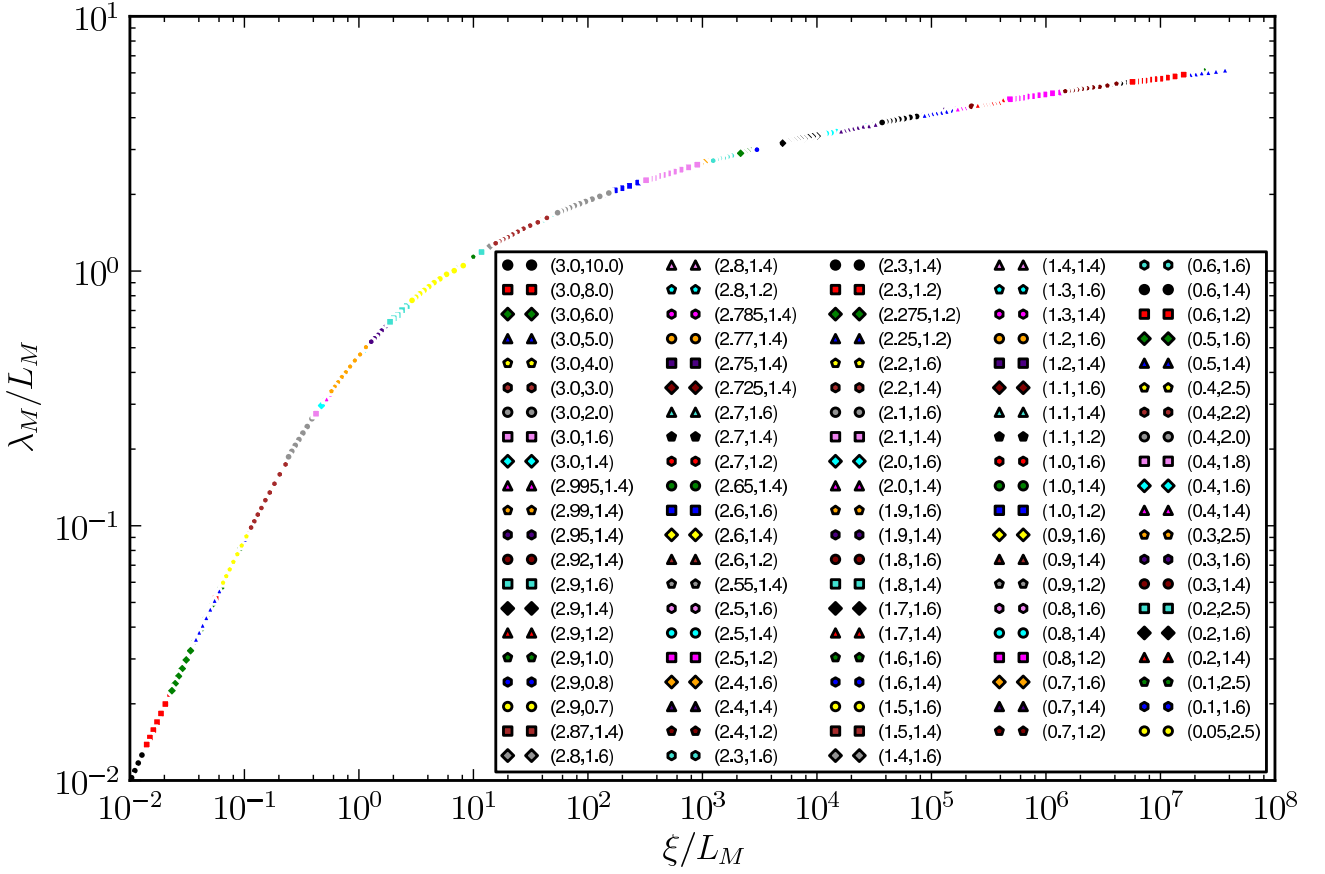}
\caption{\label{fig:collapseCurve}
(Color online) Single-parameter scaling law for the AC honeycomb lattice
with uncorrelated on-site disorder. We numerically compute the finite-size
localization length $\lambda_M$ at various energy $0.1J\leq E\leq 3J$ and
disorder strength $1.2J\leq W\leq 10J$. The fact that all data sets can be collapsed onto a single log-log
curve where $\xi$ is the infinite-lattice localization length confirms the
scaling theory of localization~\cite{abrahams1979}. Each disjoint color
represents a parameter pair ($E$, $W$) as shown in the inset.}
\end{figure*}

Based on ideas of the renormalization group~\cite{mkk1983} and the scaling
theory
of localization~\cite{abrahams1979},
it was conjectured that all data points in a $\lambda_M/L_M$ versus $\xi/L_M$
plot should
collapse onto the same universal curve,
\begin{equation}
\frac{\lambda_M}{L_M} = F \left(\frac{\xi}{L_M}\right)
\end{equation}
where the infinite-lattice localization length $\xi$ depends only on energy
$E$ and
disorder strength $W$. Our results in Fig.~\ref{fig:collapseCurve} fully
supports
this conjecture for sufficiently large $L_M$.

When $L_M \gg \xi$, the system becomes insensitive to the vertical boundary
conditions and one expects $\lambda_M\approx \xi$. The scaling function
should thus satisfy $F(x)\approx x$ for small $x$. We chose the honeycomb
AC configuration at $E=3J$ and $W=10J$ as the reference data set for this
limiting behavior, therefore fixing $\xi=(2.46\pm0.01)a$ in this case, see
black data at the bottom left corner in Fig.~\ref{fig:collapseCurve}. At
finite system width, we nevertheless expect the computed localization
length $\lambda_M$ to be slightly shorter than the true infinite-lattice
localization length $\xi$. This behavior is accounted for by a Taylor
expansion of the scaling function $F(x)=x-\alpha x^2+O(x^3).$ From our data
we infer $\alpha=(1.11\pm0.01)$.

Starting from the situation at small $x$, the full scaling function $F(x)$
is then built by finding, for each of our data sets, the corresponding
$\xi$ allowing to patch them on a single smooth curve. This procedure has
been done with our AC data sets obtained in the range $E\in[0.1J,3J]$ and
$W\in[1.2J,10J]$. We have also checked that AC data sets obtained for $E$
slightly larger than $3J$ and various $W$, as well as ZZ data sets obtained
at different energies ($E=0.4J, J,2.9J$ ) and $W=1.6J$, all collapse onto
this very same curve too. In particular, we numerically confirm that at
$E=0.4J$, $E=2.9J$, and $W=1.6J$, we get the same $\xi$ for the ZZ and AC
configurations. This might be understood as a consequence of the isotropic
dispersion relation in the two energy ranges, see Eqs.~\eqref{eq:enDisperse1}
and \eqref{eq:enDisperse2}. On the other hand, at $E=J$, the extracted $\xi$
are slightly different for the two honeycomb configurations.

\begin{figure}
	\includegraphics[width=0.48\textwidth]{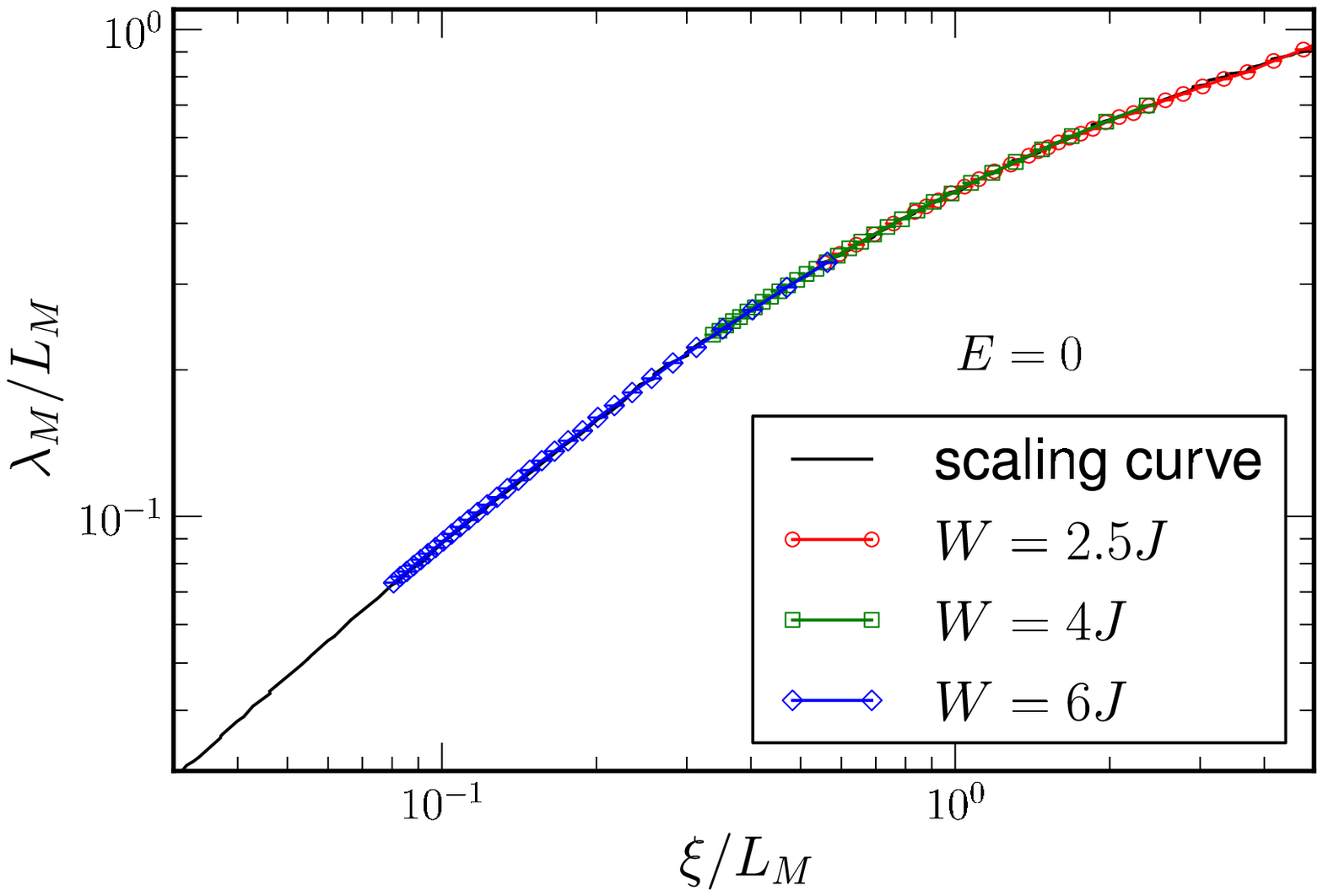}
\caption{\label{fig:xiDirac}
(Color online) Collapsing AC data obtained at $E=0$ and disorder
strengths $W\geq 2.5J$ onto the scaling function $F(x)$ shown in
Fig.~\ref{fig:collapseCurve}. For $W=2.5J$ (circles), $4J$ (squares) and $6J$
(diamonds), we have $\xi=500a$, $70.9a$ and $16.9a$ respectively. The largest
transverse sizes are given by $M=300$, $70$, and $70$ for $W=2.5J$, $4J$
and $6J$ respectively.
Data for $W=2.5J$ display a small oscillation around the universal curve.}
\end{figure}

For the AC configuration, an important finding is that data sets obtained
at the charge neutrality point ($E=0$) for $W\geq2.5J$ can all be scaled
by $F(x)$.  Fig.~\ref{fig:xiDirac} shows the collapsed data for $W=2.5J$,
$4J$ and $6J$ ($\xi=500a,70.9a$ and $16.9a$ respectively). Performing similar calculations for different lattice models (honeycomb,
square and triangular), Schreiber and Ottomeier also found that their data
obtained at $E=0$ and $W\geq4J$ could be collapsed onto a single universal
curve~\cite{schreiber1992}. Since our data at $W=4J$ and $6J$ with $M=46$
agree with theirs at $M=40$ (after proper unit conversion), we infer that
we indeed found the same universal curve.

While it is difficult to numerically confirm that data at $E=0$ and $W<2.5J$
can be scaled by $F(x)$ (see analysis below for the reason of this difficulty),
there is no apparent reason why they should not. Therefore, our results,
combined with those in Ref.~\cite{schreiber1992}, imply that all curves
$\lambda_M/L_M$ as a function of $1/L_M$ can be collapsed onto the same
universal curve by an appropriate one-parameter scaling, independently of the
lattice type and disorder strength, at least for uncorrelated box-distributed
on-site disorder and energies within (and slightly outside of) the energy
band of the clean system. This is in marked contrast with the findings of
Ref.~\cite{xiong2007}, which claim a different scaling behavior at $E=0$
(see below for a possible explanation of this apparent contradiction).

\begin{figure}
	\includegraphics[width=0.48\textwidth]{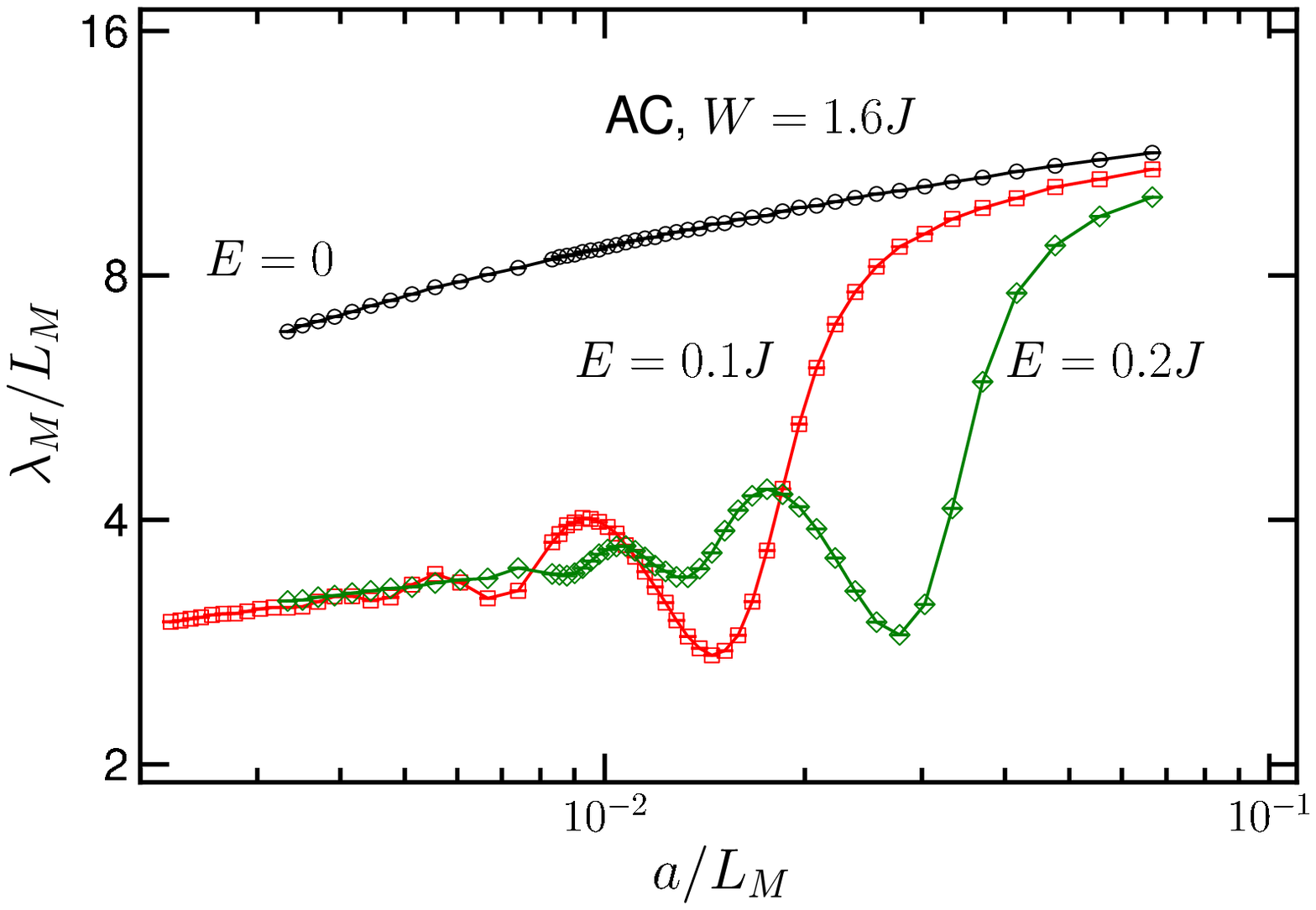}
\caption{\label{fig:loclen_weakdis}
(Color online) Plot of $\lambda_M/L_M$ as a function of $a/L_M$ at weak
disorder $W=1.6J$. The oscillations observed at small $L_M$ results from
the opening of new scattering channels
when $L_M$ is gradually increased. In the linear dispersion regime, the
oscillation period in $L_M$ is crudely approximated as inversely proportional
to $E$, with a small correction due to disorder.}
\end{figure}

For the clean system near $E=0$, the energy shell slices the band structure
near the Dirac points, defining two circles of allowed wavevectors centred
on $\vec{K}$ and $-\vec{K}$. The initial state $|\vec{k}s\rangle$ belongs to
one of these circles, say the one around $\vec{K}$. Introducing the deviation
wavevector $\vec{q}=\vec{k}-\vec{K}$, the circles have radius $|\vec{q}|$. In
the presence of disorder, the two circles are broadened in the energy shell and
form rings of allowed wavevectors with mean radius $|\vec{q}|$. Since disorder
is uncorrelated in space, the two rings are coupled by scattering. However,
the scattering processes being elastic, starting from the initial state
$\vec{q}$, only these two rings around the Dirac points will be populated
in the course of time. Loosely speaking, if scattering does change the
direction of $\vec{q}$, it cannot change its modulus. Around $E=0$, the
dynamics is thus characterized by small wavevectors, i.e. by long distances
in real space. This is why we expect  finite-size effects to be larger around
$E=0$. This can be seen in Fig.~\ref{fig:loclen_weakdis} where the dependence
of $\lambda_M$ on $L_M$ shows a damped oscillatory behavior when $L_M/a$
is not large enough. As explained below, the very existence of oscillations
can be traced back to the opening of new scattering channels when $L_M$
increases whereas the damping can be traced back to disorder broadening.

Consider the finite-size AC configuration. For periodic boundary conditions,
the allowed values for the wavevector $\vec{k}$ along $Ox$ and $Oy$ are
quantized according to $k_x= n_x \Delta k_N$ and $k_y = n_y \Delta k_M$
with $\Delta k_N = 2\pi/L_N$, $\Delta k_M = 2\pi/L_M$, $n_x \in \{0, \cdots,
N\}$ and $n_y \in \{0, \cdots, M\}$. In the vicinity of $E=0$, the curve at
constant $E$ will enclose a small circular area around the Dirac point $K$ in
the Brillouin zone containing few such discrete points. When the system size,
notably $L_M,$ is increased, more points enter this area, which means that
more channels open for propagation. The period of the oscillations observed in
Fig.~\ref{fig:loclen_weakdis} thus corresponds to the change in $L_M$ allowing
one new channel to open, i.e.  $\Delta L_M=2\pi/q,$ where $q$ is the radius
of the circle at energy $E,$. Using Eq.~\eqref{eq:enDisperse2}, we estimate
the period of the oscillations to be roughly of the order of $\Delta L_M=3\pi
aJ/|E|.$  However, each quantized $\vec{k}$-mode corresponding to energy $E$
is broadened by disorder, typically by $\ell^{-1}$. As a consequence modes
cannot be distinguished from each other when $\Delta k_M \ell \approx 1$. Near
the Dirac point, one has $\ell = c \tau$ and we find that the oscillations
due to the opening of new channels are washed out as soon as
\begin{equation}
\frac{a}{L_M} \lesssim - \frac{2{\rm Im}\Sigma(E,W)}{3\pi J}.
\end{equation}
Using SCBA at $W=1.6J$, this simple estimate gives $a/L_M \approx 4 \times
10^{-3}$ at $E=0.1J$ and $a/L_M \approx 7.4 \times 10^{-3}$ at $E=0.2J$,
in perfect agreement with the data collected in Fig.~\ref{fig:loclen_weakdis}.

To obtain a reliable estimate of $\xi$, it is necessary to go beyond the
oscillatory region. This is numerically challenging at $E=0$ for weak
disorder, see \eqref{eq:scba0}. For example, the SCBA estimate at $W=1.6J$
now gives $a/L_M \approx 2 \times 10^{-6}$, way out of the capabilities of the
RGF scheme. We believe this is the reason why a different scaling at $E=0$
was claimed in \cite{xiong2007}: the sample sizes used by the Authors were
probably not large enough to reach the scaling region. As seen in Fig. \ref{fig:loclen_weakdis}, it is easy to
miss the oscillations at $E=0$ for small $W.$
The curve there is clearly different from the other ones and cannot be
collapsed by translation (in log scale) onto the universal curve, $F(x).$ This
might be the origin of the claim in Ref.~\cite{xiong2007} that a different
scaling is observed at $E=0.$ This is however only a finite-size effect.

\begin{figure}
	\includegraphics[width=0.48\textwidth]{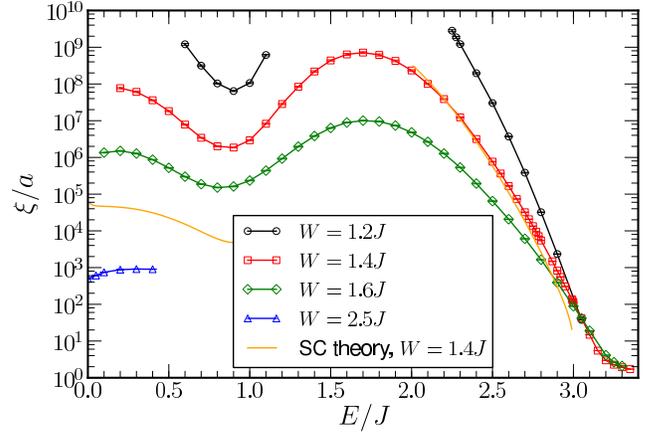}
\caption{\label{fig:xi}
(Color online) Infinite-lattice localization length $\xi$ (in units of $a$)
versus energy $E$ (in units of $J$) for various disorder strengths $W$. For
each set ($E,W$), $\xi$ is extracted from our numerically-computed $\lambda_M$,
Eq.~\eqref{eq:loclen}, with the help of the scaling function $F(x)$ shown in
Fig.~\ref{fig:collapseCurve}. The orange solid line gives $\xi=2\ell \sqrt{\Exp{\pi \kappa\ell}-1}$
at $W=1.4J$ as inferred from the self-consistent theory of localization,
see Eqs.~\eqref{eq:kfdefSCTL}.
In this analytical prediction, the scattering mean free path $\ell$ is
estimated using SCBA and $\kappa$ is defined in Eq.~\eqref{eq:kfdefSCTL}. If
the qualitative behavior is relatively satisfactory, the quantitative
predictions are off by several orders of magnitude in the linear dispersion
regime and by about an order of magnitude in the quadratic dispersion regime.}
\end{figure}

In Fig.~\ref{fig:xi}, we show the variations of the localization length $\xi$
as a function of the energy $E$ for different disorder strengths $W$. In the
linear dispersion regime (but not too close to $E=0$), $\xi$ appears to be
a constant whose magnitude increases as $W$ decreases. When getting closer
to $E=0$, $\xi$ decreases (see curve at $W=2.5J$) and a small dip occurs. A
similar feature is predicted in Ref.~\cite{lherbier2008} when computing $\xi$
from the numerically-evaluated semi-classical conductivity. Extrapolating our
results, we find that $\xi$ is finite at $E=0$ but increases sharply when $W$
decreases. Taking graphene as an example ($a=0.142{\rm nm}, J\approx 2.7{\rm
eV}$), the localization length $\xi$ is of the order of $10^6a\approx 0.1{\rm
mm}$ at $W=1.6J \approx 4.32{\rm eV}$. This sample size is achievable with
the current state-of-the-art technology.

A second notable feature is the small dip slightly below the van Hove
singularity at $E=J$. This is surprising as the DoS at the van Hove
singularity diverges in the absence of disorder, which implies more propagation
channels. Finally we see that $\xi$ decreases as $E$ gets closer to the band
edge ($E=3J$) and slightly extends beyond it.

To conclude this subsection, we compare in Fig.~\ref{fig:xi} our numerical
data near $E=0$ and $E=3J$ at $W=1.4J$ with the analytical prediction of the
self-consistent theory of localization	\cite{vollhardt1980} (see section
\ref{sec:sctl}). As can be seen, if the general trend is qualitatively
satisfactory and even semi-quantitatively correct (within a factor 10) near
the band edge, the prediction is off by more than 3 orders of magnitude near
the band centre $E=0$.

\subsection{Scattering mean free path $\ell$}\label{sec:mfp}

\subsubsection{RGF numerical estimate}

The elastic scattering mean free path $\ell$ defines the distance a particle
travels on average without being scattered. For the negative energy sector
(see the reason why in the following subsection), we compute the 1D averaged
retarded wave function as
\begin{equation}
	\overline{\Psi}_{N} = \frac{J}{M} \overline{\sum_{m,m'=1}^M
	\bigg[G^{(N+2n)}_{n, n+N}\bigg]_{m,m'}},
\label{eq:1dwavefunc}
\end{equation}
where the retarded submatrix $G^{(N+2n)}_{n, n+N}$ connecting slices $n$
and $(n+N)$ is evaluated at complex energy $E+\mri \eta$ ($E \leq 0$) and
disorder strength $W$, $\eta$ being a small positive number. The indices $m$
and $m'$ label the sites within each slice respectively. In the following, $n$
is chosen such that the distance between slice $n$ and the closest open-end
boundary is large enough (see below).

\begin{figure}
	\includegraphics[width=0.4\textwidth]{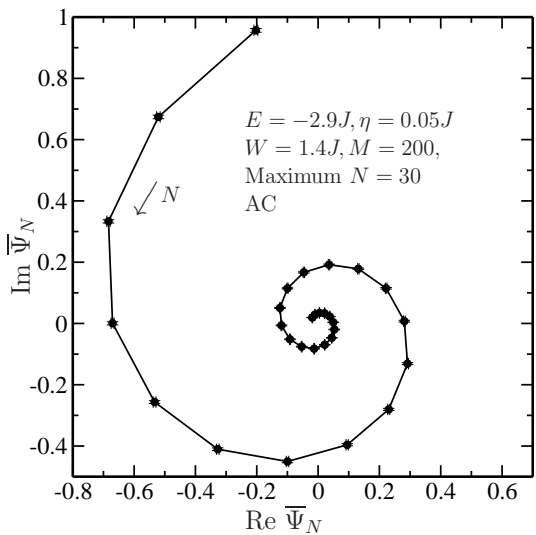}
\caption{\label{fig:spiralGreen}
Disorder-averaged wave function $\overline{\psi}_N$ shown in the complex
plane as the longitudinal size of the lattice, fixed by the number of
slices $N$, increases (see arrow for the flow). The plot has been made for the
honeycomb AC configuration with $M=200$ sites per slice. The complex energy
is $E+\mri\eta=-2.9J+\mri0.05J$ and the disorder strength is $W=1.4J$. The
spiraling feature shows that
$\overline{\Psi}_N \sim \exp\big({\mri k_w(\eta)
L_N}\big)\times\exp\left({-\frac{L_N}{2\ell(\eta)}}\right)$, from which we
can extract the decay constant $\ell(\eta)$. The scattering mean free path
is further obtained as $\lim_{\eta\to0}\ell(\eta)$.}
\end{figure}

As shown in Fig.~\ref{fig:spiralGreen}, $|\overline{\Psi}_{N}|^2$ falls
off exponentially with a decay constant $\ell(\eta)$. The scattering mean
free path is further obtained as $\ell =\lim_{\eta\to0}\ell(\eta)$. In our
simulations, the number of disorder configurations has been chosen such
that we have $2\times10^5$ samples near the band edge at $-3J$ (quadratic
dispersion regime) and $2\times10^6$ samples near the band centre (linear
dispersion regime). We have imposed a minimum distance of $5\ell(\eta)$
between slice $n$ and its closest open-end boundary, thereby fixing the value
of $n$ in Eq.~\eqref{eq:1dwavefunc}. This ensures that the exponential decay
is insensitive to the open-end boundaries. Our numerical
results are thus equivalent to those that would have been obtained with
an infinite tube. We would like to mention here an important technical
remark. When summing the submatrix entries in~\eqref{eq:1dwavefunc}, serious
numerical cancellation errors occur when $E$ and $J$ have same signs. In
our case $J>0$ and the numerical evaluation is stable only if $E<0$. The
reason behind this numerical instability is the relative sign between the
quasi-Bloch sublattice components of $|\vec{k}s\rangle$, see Eq.~\ref{eq:ks}.

\begin{figure}
	\includegraphics[width=0.48\textwidth]{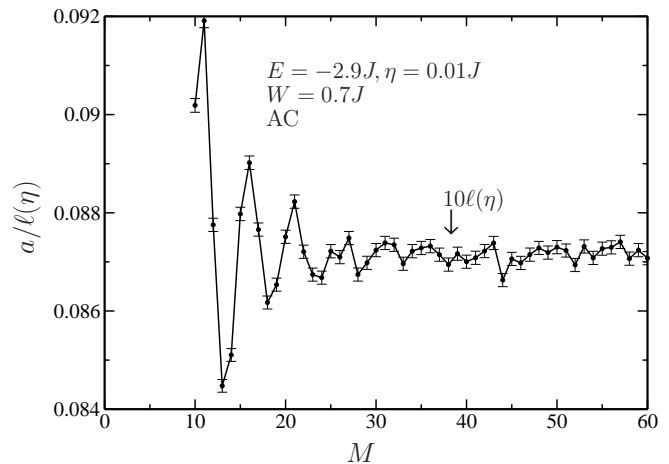}\caption{\label{fig:oscillatingls}Oscillation
	of $1/\ell(\eta)$ as a function of the transverse size $M$. The
	oscillation period can be accounted for by the presence of new
	scattering channels. The oscillation stops when the system does not
	sense the periodic boundary condition, i.e. $L_M>10\ell(\eta)$.}
\end{figure}

\begin{figure}
	\includegraphics[width=0.48\textwidth]{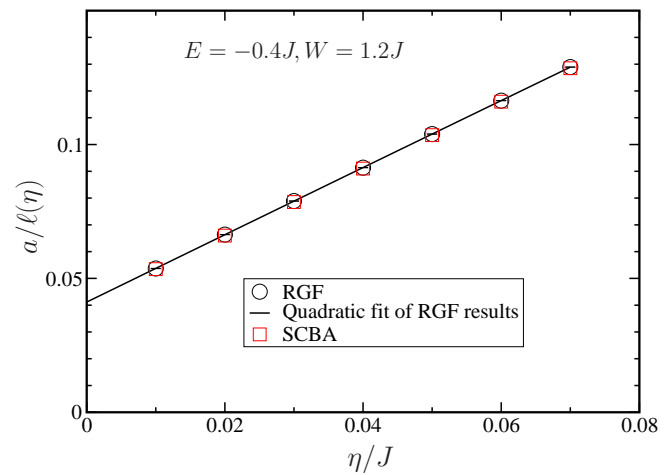}
\caption{\label{fig:lseta}
(Color online) Evaluation of the infinite-lattice scattering mean free
path $\ell$ at a fixed energy $E$ and disorder strength $W$ ($E=-0.4J$ and
$W=1.2J$ in the figure). A sequence of reliable estimates $a/\ell(\eta)$
is first obtained by the RGF technique for smaller and smaller $\eta$ (black
open circles). A quadratic fit (solid black line) is then used to interpolate
the value at $\eta=0$, from which $\ell$ is deduced. For comparison, we give
the results obtained with the SCBA method (open red squares). For the chosen
disorder strength the agreement is excellent.}
\end{figure}

The decay constant $\ell(\eta)$ is extracted by performing a linear fit on
$\ln |\Psi_N|$ as a function of $L_N$. We see in Fig.~\ref{fig:oscillatingls}
that $a/\ell(\eta)$ displays oscillations similar to those observed for $\xi$
when the transverse size (fixed by the number of sites $M$) increases.
Again, the oscillations can be accounted for by the opening of new
scattering channels. By further increasing $M$, the oscillations damp
and finally the fluctuations in $a/\ell(\eta)$ reach the level of the
error bars themselves. This happens when the width $L_M$ roughly exceeds
$10\ell(\eta)$. Beyond this point, the system does not feel anymore the
transverse periodic boundary condition and the corresponding $\ell(\eta)$
are reliable estimates of the infinite-lattice case. We show how to extract
$\ell$ from these reliable estimates in Fig.~\ref{fig:lseta}. Note that
a direct evaluation of $\ell$ at $\eta=0$ is in principle possible, but
is unfortunately affected by huge fluctuations as the Green's function is
computed on the real axis where its poles lie. This direct method proves
thus much less efficient in practice.

\subsubsection{SCBA estimate}

In the same spirit, an SCBA estimate of $\ell$ can be obtained by using
$G_0(E-\Sigma(E))$ in \eqref{eq:1dwavefunc} and by considering a finite
horizontal lattice strip of length $L$ with periodic boundary conditions in
the transverse direction in the limit $L_M, L \to \infty$. In the limit of
an infinite AC honeycomb lattice, a careful but straightforward calculation
using Eqs.~\eqref{eq:ks} and \eqref{eq:G0} leads to

\begin{equation}
\label{eq:Gscba}
\overline{\Psi}_N = J \ \int_0^{2\pi} \ \frac{dq}{2\pi} \
\frac{e^{\mri Nq}}{z+J (1+2\cos q)},
\end{equation}
where $z= E-\Sigma(E)$ and $q = \sqrt{3}k_xa/2$. The self-energy $\Sigma(E)$
is computed using Eqs.~\eqref{eq:scba1} and \eqref{eq:Izfull}. The distance
between the two slices being $L_N=\sqrt{3}aN/2$ in the AC configuration,
we then get the scattering mean free path as
\begin{equation}
\frac{1}{\ell} = - \lim_{N\to \infty} \frac{\ln |\overline{\Psi}_N|^2}{L_N}.
\end{equation}
The exponential decay of $\overline{\Psi}_N$ is given by the imaginary part
of the complex pole $Q=Q_r+\mri Q_i$ solving
\begin{equation}
	\cos Q = -\frac{1+z/J}{2}
	\label{eq:zmomentum}
\end{equation}
with $Q_i \geq 0$. The scattering mean free path is then just
$a/\ell=4Q_i/\sqrt{3}$. Approximate solutions are given in the
Appendix~\ref{AppendixSCBA}.

A word of caution is here necessary. Inspection of \eqref{eq:Gscba} in the
absence of disorder ($\Sigma =0$) shows that the allowed energy range is
restricted to $-3J \leq E \leq J$. A part of the positive energy sector is
thus missed and will keep missed at weak enough disorder. As a consequence
using Eq.~\eqref{eq:Gscba} is only well adapted to the negative energy sector.
The physical reason is that the honeycomb lattice has a two-point Bravais
cell and the prescription \eqref{eq:1dwavefunc} amounts to consider a
symmetric combination of amplitudes associated to \textsc{a}-sites and
\textsc{b}-sites within a Bravais cell of the initial slice. Would one had
chosen the antisymmetric combination, then one would have gotten
\begin{equation}
\overline{\Psi}_N = J \int_0^{2\pi} \ \frac{dq}{2\pi} \
\frac{e^{\mri Nq}}{z-J (1+2\cos q)},
\end{equation}
which is well adapted to the positive energy sector. As a rule of thumb,
we thus only use Eqs.~\eqref{eq:1dwavefunc} and \eqref{eq:Gscba} for the
negative energy sector $E=-|E|$ and then resort to \eqref{eq:propsig}
whenever necessary.

Once the scattering mean free path $\ell(E)$ and the scattering mean free
time $\tau(E)$ have been calculated, one can compute the ratio $v(E) =
\ell(E)/\tau(E)$. From \eqref{eq:zmomentum}, we get
\begin{equation}
\frac{v(E)}{c} = \frac{2}{\sqrt{3}} \ \sin Q_r \ \frac{\sinh Q_i}{Q_i},
\end{equation}
where $c$ is the Dirac fermions speed and where $0 \leq Q_r \leq 2\pi/3$ for
propagation from left to right. At sufficiently weak disorder, one expects $Q_i
\ll 1$, and thus $v(E)/c \approx (2 \sin Q_r )/\sqrt{3}$, where $Q_r$ solves
\begin{equation}
\varepsilon(Q_r) = -J(1+2\cos Q_r) \approx E - {\rm Re}\Sigma(E).
\end{equation}
In the weak disorder regime, the real part of self-energy is generally small
compared to the value of $E$ and can be usually discarded.

\begin{figure}
	\includegraphics[width=0.48\textwidth]{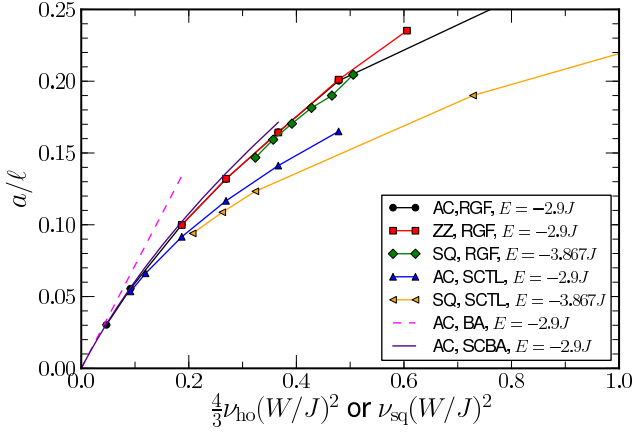}
\caption{\label{fig:ls_E-2.9}
(Color online) Inverse of the scattering mean free path $\ell$ (in units of the inverse of lattice
constant $a$) extracted from the recursive Green's function (RGF) method
and from the self-consistent Born approximation (SCBA) as a function of the
square of the disorder strength $W$ in units of the hopping energy $J$. To
compare results obtained for the honeycomb and square lattices, $(W/J)^2$
is renormalized by a factor proportional to the corresponding densities of
states, $\nu_{\rm ho}=0.14$ (honeycomb) and $\nu_{\rm sq}=0.081$ (square),
in the absence of disorder as defined by Eq.~\eqref{eq:dosDef}. For both
lattices, the (negative) energy $E$ is chosen near the band edge where
the dispersion is quadratic. The inset identifies the different lattices,
configurations, energies and calculation methods.
The overlap between the results of the armchair (AC) honeycomb and square (SQ)
lattices show that the average propagation near the band edge is independent
of the lattice type and solely determined by the quadratic nature of the
dispersion relation.  Comparison with the Born approximation is shown.
We also plot the scattering mean free path corresponding to the numerically
observed localization length (calculated with the Recursive Green Function method),
assuming that the two quantities are connected by eq.~(\ref{eq:kfdefSCTL}) derived
from the Self-Consistent Theory of Localization.
This shows that the predictions of the Self-Consistent Theory of Localization
are qualitatively correct, especially for weak disorder, but that
large quantitative deviations are observed at large disorder.
}
\end{figure}

\begin{figure}
	\includegraphics[width=0.49\textwidth]{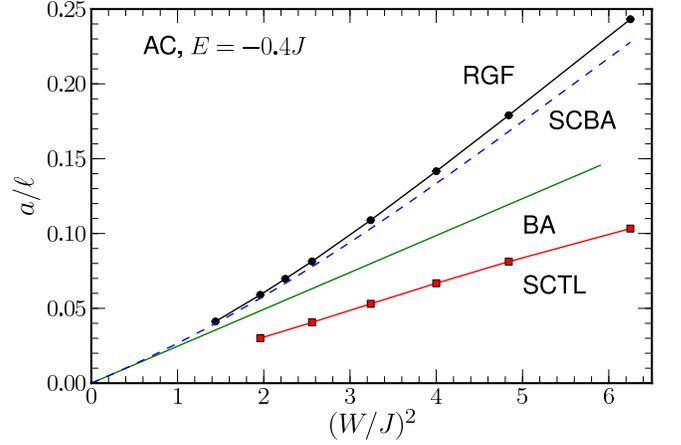}
\caption{\label{fig:ls_E-0.4}
(Color online)  Inverse of the scattering mean free path $\ell$ (in unit of the inverse of the lattice
constant $a$) as a function of the square of the disorder strength $W$ in
units of the hopping energy $J$ for the AC honeycomb lattice. The (negative)
energy $E=-0.4J$ is chosen in the linear dispersion regime. Connected filled
black circles: RGF results. 
Blue dashed line:
SCBA results. Black solid line: Born approximation (BA). 
Connected filled red squares: scattering mean free path estimated from the
localization length, assuming that they are connected by eq.~(\ref{eq:kfdefSCTL}) derived
from the Self-Consistent Theory of Localization.
While the SCBA prediction agrees well with the RGF numerical computation
of the mean free path, the predictions of the Self-Consistent Theory of Localization
are quantitatively off. 
}
\end{figure}

\begin{figure}
	\includegraphics[width=0.48\textwidth]{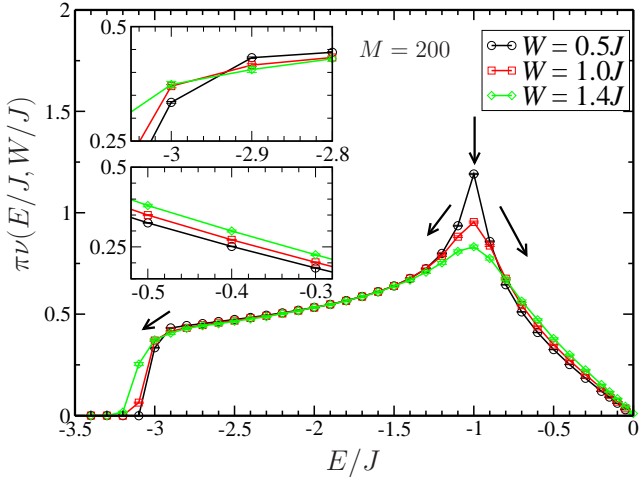}
\caption{\label{fig:dos_full_compare}
(Color online) Density of states (DoS) $\nu(E/J,W/J)$, Eq.~\eqref{eq:dosdef}
as a function of (negative) energy $E$ (in units of the hopping energy $J$)
for several disorder strengths $W$. The number of sites in the transverse
direction is $M=200$. The DoS is increasingly smoothed by disorder as $W$
is increased. Because of particle number conservation, the area under the
curve is also a conserved quantity and states are simply redistributed over
a broader energy range. For example, as the smoothed van Hove singularity
peak at $E=-J$ decreases, the corresponding states are redistributed in the
wings as indicated by the arrows. This translates into an increased DoS
near the charge neutrality point $E=0$. Similarly, states at the band edge at
$E=-3J$ are partly redistributed outside the energy band of the clean system
($E<-3J$), resulting in a decrease of the DoS in the quadratic dispersion
regime $E\gtrsim -3J$.
The upper inset shows the variations of the DoS in the quadratic dispersion
regime while
the lower inset shows the variations of the DoS in the linear dispersion
regime near the charge neutrality point.}
\end{figure}

\begin{figure}
	\includegraphics[width=0.48\textwidth]{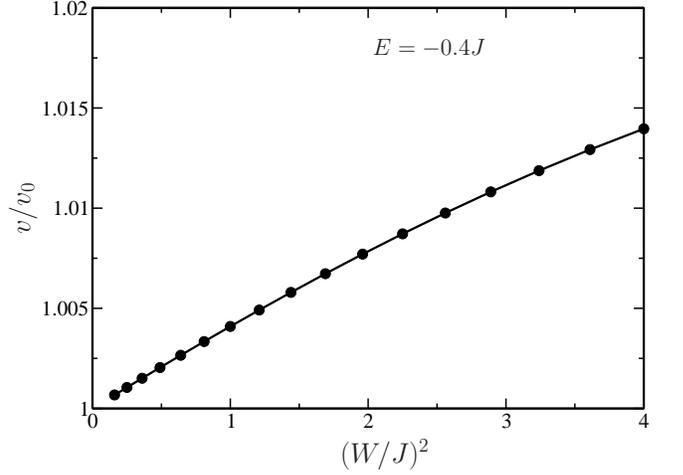}
\caption{\label{fig:v_E-0.4}
Variation of the group velocity $v=\ell/\tau$ with respect to the square of
the disorder
strength $W$ in units of of the hopping energy $J$ within the SCBA. The
(negative) energy $E=-0.4J$ is chosen near the linear dispersion regime. The
group velocity of the clean system (chosen along $Ox$) is $v_0=1.1
c$, where $c$ is the massless Dirac particles velocity. As one can see,
the group velocity near the charge neutrality point is only slightly
enhanced by disorder. The reason why SCBA overestimates $\ell$ near the
charge neutrality point compared to the RGF calculation is thus essentially
a consequence of the smoothing of the density of states by disorder, see
Fig.~\ref{fig:dos_full_compare}.}
\end{figure}

\begin{figure}
	\includegraphics[width=0.48\textwidth]{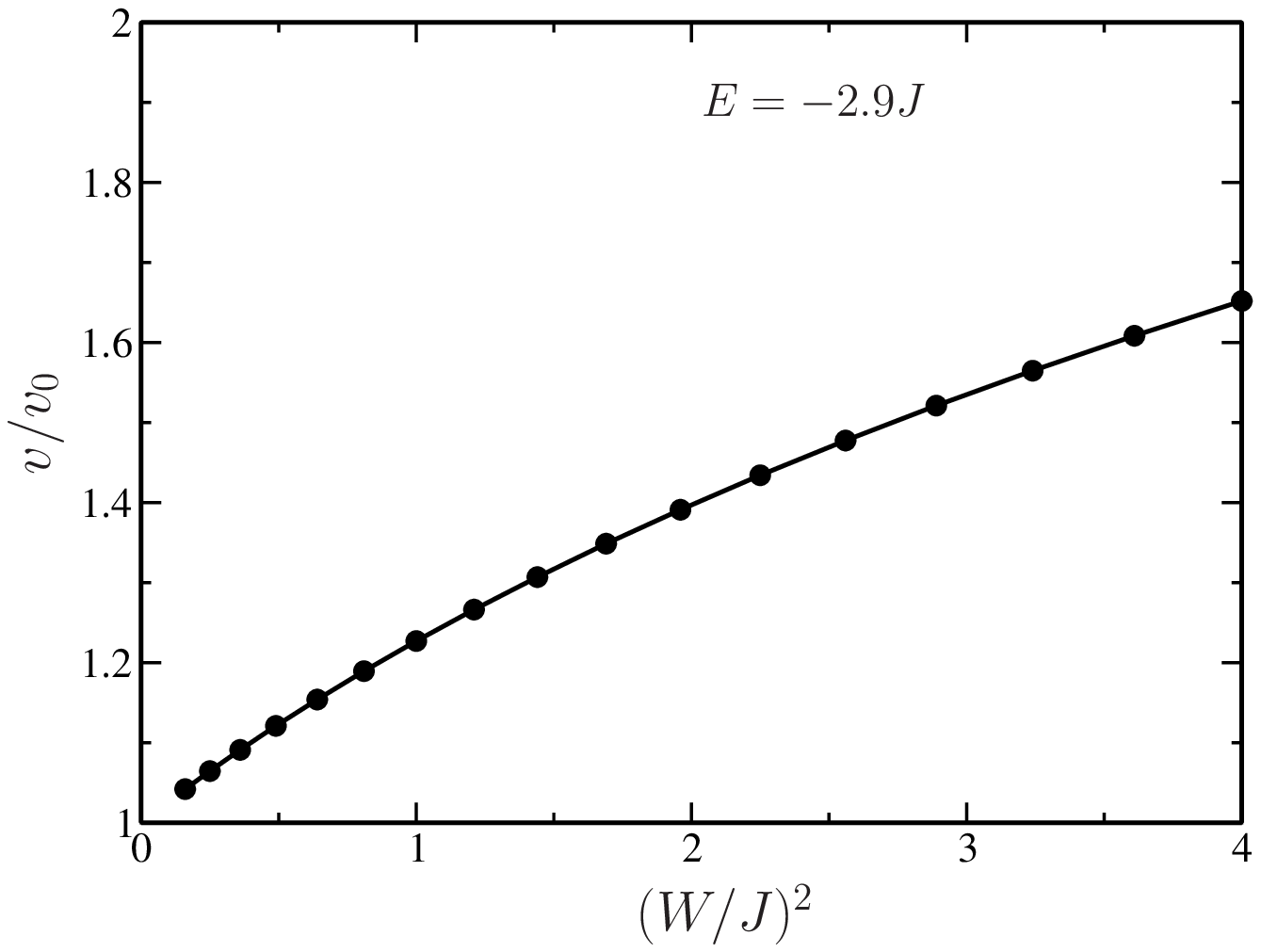}
\caption{\label{fig:v_E-2.9}Variation of the group velocity $v=\ell/\tau$ with respect to the square of
the disorder
strength $W$ in units of of the hopping energy $J$ within the SCBA. The
(negative) energy $E=-2.9J$ is chosen in the quadratic dispersion regime. The
group velocity of the clean system (chosen along $Ox$) is given by $v_0 =
0.36 c$, where $c$ is the massless Dirac particles velocity. As one can
see, the group velocity near the band edge is significantly enhanced by
disorder. This is the main reason why SCBA underestimates $\ell$ near the
band edge compared to the RGF calculation. }
\end{figure}

Returning to fully dimensioned quantities, we thus find the usual result
that $v(E) = |\partial_{k_x} \varepsilon(k_x)|/\hbar$ is the group velocity
(here along $Ox$) when disorder is sufficiently weak.
Quantitative comparison between the SCBA and RGF results are shown
in Figs.~\ref{fig:ls_E-2.9} and \ref{fig:ls_E-0.4}. The two methods show
remarkable agreement. We note however that the SCBA underestimates $\ell$
in the quadratic dispersion regime
but overestimates $\ell$ in the linear dispersion regime. The deviation of
the Born approximation from the true (RGF) $\ell$ might be understood from
two perspectives: first as a consequence of the smoothing of the DoS $\nu(E)$
by disorder (see Fig.~\ref{fig:dos_full_compare}), second
as a consequence of the enhancement of the group velocity (see
Fig.~\ref{fig:v_E-0.4} and \ref{fig:v_E-2.9}). Indeed, from the relation
$\ell=v\tau$ ($v$ is the group velocity), we know that $1/\ell$ is directly
proportional to ${\rm Im}\Sigma(E)$, which is in turn directly proportional
to the DoS. As disorder is increased, the sharp variations of the DoS at
the band edges ($E=\pm 3J$), at
the van Hove singularities ($E=\pm J$), and at the charge neutrality point
($E=0$) are smoothed.
Since the DoS area must be conserved (particle number is conserved), this
smoothing is accompanied by a redistribution of states over energies and
by an increase of the DoS at some energies. For example, since the van Hove
singularity peaks in the DoS are decreased, there is a corresponding increase
of the DoS in the linear dispersion regime. Similarly, as the band edges are
smoothed, the DoS at $|E|<3J$ decreases but increases for $|E|>3J$. Therefore,
the departure of the actual $1/\ell$ from the Born value as disorder is
increased, and whether the Born value underestimates or overestimates the
actual $1/\ell$ in Fig.~\ref{fig:ls_E-2.9} and Fig.~\ref{fig:ls_E-0.4},
is a simple consequence of the smoothing effect of the DoS in different
energy regimes.

\subsubsection{Comparison with the self-consistent theory of localization}
\label{sec:sctl}
The self-consistent theory of localization
(SCTL)~\cite{vollhardt1980,*wolfle2010} provides a self-consistent recipe
which extends the weak-disorder diagrammatic approach to the regime of
Anderson localization. The diagrammatic approach describes perturbatively the weak localization corrections to classical transport
due to interference along closed loops. As this correction diverges for infinite system size,
the self-consistent theory of localization aims at computing the minimum size of the system
beyond which the correction is strong enough to stop the diffusive transport and identifies
this length scale with the localization length.

For isotropic scattering and an isotropic dispersion relation,
SCTL establishes a simple link between the localization length $\xi$, the
scattering mean free path $\ell$ and the wavevector $\kappa$,
\begin{equation}
\label{eq:kfdefSCTL}
\xi =  2 \ell \ \sqrt{\Exp{\pi\kappa \ell}-1} \approx 2\ell \Exp{\frac{\pi}{2}\kappa \ell},
\end{equation}
where $\kappa = |\vec{k}|$ when the energy is chosen near the band edges
and $\kappa = |\vec{q}|=|\vec{k}-\vec{K}|$ when the energy is chosen near
the charge neutrality point. SCLT being valid when $\kappa\ell\gg 1$,
we further get the approximation
\begin{equation}
\frac{\pi\kappa\ell}{2} \approx \ln(\pi \kappa\xi/4)-\ln\bigl(\ln(\pi \kappa\xi/4)\bigr).\label{eqn:sctl}
\end{equation}
We show this SCTL prediction in Fig.~\ref{fig:ls_E-2.9} and
Fig.~\ref{fig:ls_E-0.4}, $\xi$ being evaluated by the RGF method. Much to our
surprise, the agreement between our numerical data and the SCTL prediction
proves very poor. It should be reminded however that SCLT is supposed to
be valid when $\kappa\ell\gg 1,$ that is in the weak disorder limit where
$a/\ell$ is simply proportional to $W^2$. Fig.~\ref{fig:ls_E-2.9} (and to a
lesser extent Fig.~\ref{fig:ls_E-0.4} too) indeed shows that the SCLT estimate
tends to be a linear function of $W^2$. In fact all predictions, SCBA, RGF,
SCTL and Born approximation agree well when $W \ll J$. It would be desirable
to have numerical results at smaller disorder strengths. Unfortunately,
the localization length is too large to be measurable. In any case, our
numerical results show that higher orders in $W$ are completely different
for the true (RGF) $a/\ell$ and the SCLT prediction. Furthermore, the trend
in the linear dispersion regime is completely different. These results are
yet to be understood.

\subsection{Density of states}
\label{SubSec:DoS}
The disorder-averaged DoS per lattice site is defined by
\begin{equation}
	\nu(E/J,W/J)=\frac{1}{2N_c}\sum_\mu
	\overline{\delta(E/J-\lambda_\mu/J)},\label{eq:dosDef}
\end{equation}
where $N_c$ is the total number of Bravais cells and where $\lambda_\mu$
is the eigenvalue of the Hamilton operator $H$ in Eq.~\eqref{eq:tightBinding}.
This definition is related to the diagonal elements of the Green's function by
\begin{equation}
\label{eq:dosdef}
	\nu(E/J,W/J)=-\frac{J}{\pi}\lim_{\eta\to0^+}{\rm Im}\langle
	j|\overline{G}(E+\mri\eta)|j\rangle,
\end{equation}
where $j$ labels an arbitrary lattice site in the infinitely-large lattice. Like what we did to extract $\ell$ with the RGF method, we compute
$\overline{G}(E+\mri\eta)$ for $\eta/J\in[0.01,0.07]$ and then perform a
quadratic fit in $\eta$ to extract the limit $\eta \to 0^+$. For each value
of $\eta$, the longitudinal length $L_N$ and the transverse width $L_M$
are both chosen
to be greater than $10 \ell(\eta)$. The value of $\langle
j|\overline{G}(E+\mri\eta)|j\rangle$ is then estimated by considering any
site $j$ at a minimum distance of $5\ell(\eta)$ from the two ends of the tube.

\begin{figure}
	\includegraphics[width=0.47\textwidth]{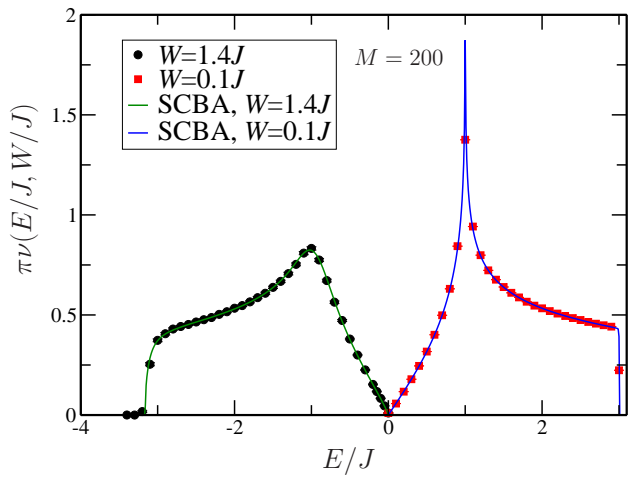}
\caption{\label{fig:dos}
(Color online) Density of states $\nu(E/J,W/J)$ computed by the recursive Green's
function method and by the self-consistent Born approximation at two weak
disorder strengths $W$. The number of sites in the transverse direction is
$M=200$. Since $\nu(E/J,W/J)$ is an even function of $E$, we have plotted the RGF
(symbols) and SCBA (continuous line) predictions at $W=1.4J$ in the negative
energy sector. The corresponding predictions at $W=0.1J$ have been plotted
in the positive energy sector. The agreement is excellent.}
\end{figure}

\begin{figure}
	\includegraphics[width=0.48\textwidth]{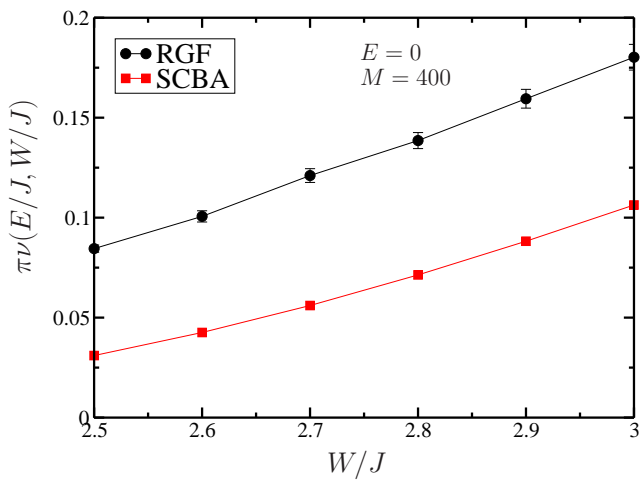}
\caption{\label{fig:dosDirac}
(Color online) Density of states $\nu(E/J,W/J)$ computed by the recursive
Green's function method (full black circles) and by the self-consistent Born
approximation (red squares) as a
function of the disorder strength $W$ (in units of the hopping energy $J$)
at the charge neutrality point $E=0$. SCBA underestimates $\nu(E/J,W/J)$ by more than a factor 2
when disorder is large enough. This is the main reason why SCBA overestimates
$\ell$ (see text).}
\end{figure}

The comparison between the DoS calculated with the RGF and SCBA methods is
shown in Fig.~\ref{fig:dos}. The two methods agree remarkably well except near
the band edges ($E=\pm 3J$) and at the charge neutrality point ($E=0$). The
breakdown of SCBA is probably due to the fact that $q\ell\ll1$. For moderate
disorder strength ($W=1.4J$), the RGF results display Lifshitz tails near
the band edges while the SCBA results show a square-root cut-off.

The DoS at $E=0$ as a function of $W$ is shown in Fig.~\ref{fig:dosDirac}. The
large deviation between the RGF and SCBA results is expected since the latter
is strictly not valid in the range of disorder strengths shown. Note that Shon
and Ando~\cite{shon1998} obtained slightly different SCBA results for the DoS
at the Dirac points probably because they used a strictly linear dispersion
relation, hence discarding trigonal warping~\cite{neto2009}. As already
witnessed by the results on $\ell$, the SCBA underestimates $\nu(E)$, and
hence the self-energy $\Sigma(E)$, in the linear dispersion regime. However,
although quantitatively incorrect, the SCBA qualitatively captures an essential
property of the system, namely the very fast decrease of the DoS when $W$
goes to zero, essentially like $\exp(-A/W^2).$

\section{Conclusion}\label{sec:conclusion}
In this paper, we have studied coherent transport in the honeycomb lattice
subjected to the effect of a spatially-uncorrelated on-site disorder with
symmetric box-like distribution. We have used the recursive Green's function
method to reliably extract the scattering mean free path $\ell$ and the density
of states. We have compared these quantities to the analytic predictions
of the self-consistent Born approximation and found good agreement at weak
disorder. We have also used the recursive Green's function method to extract
the localization lengths for different transverse sizes. We have shown that
all of these finite-size localization lengths can be collapsed onto a single
curve. We have checked that this curve is universal as it applies equally
well to the square and honeycomb lattices at any energy. In particular,
it applies to the honeycomb lattice at the charge neutrality point, at the
van Hove singularities, and at the band edges. These findings validate the
one-parameter scaling hypothesis which is thus not restricted to particles
with either quadratic dispersion or linear dispersion relations.

\begin{acknowledgments}
LKL acknowledges support from the French Merlion-PhD programme (CNOUS
20074539) and would like to thank M. Schreiber for sharing of data. BG and ChM
acknowledge support from the LIA FSQL and from the France-Singapore Merlion
programme (FermiCold grant 2.01.09). The Centre for Quantum Technologies is
a Research Centre of Excellence funded by the Ministry of Education and the
National Research Foundation of Singapore. ChM is a Fellow at the Institute
of Advanced Studies (NTU, Singapore).
\end{acknowledgments}

\appendix

\section{Diagonal elements of the clean lattice Green's function}
\label{AppendixG0}
\begin{figure}[!ht]
	\includegraphics[width=0.45\textwidth]{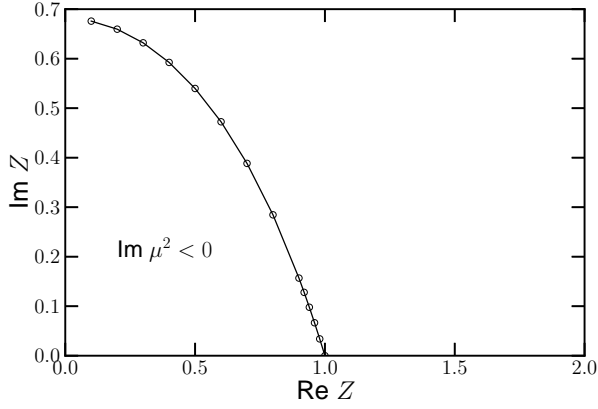}
\caption{\label{fig:scba_m2_boundary} The boundary ${\rm Im}\mu^2=0$
(open circles) divides the first quadrant of the complex-$Z$ plane in two
regions. The left region satisfies ${\rm Re}Z < 1$ and ${\rm Im}\mu^2 < 0$.
Because of the analytical continuation of the elliptic integral across the
boundary ${\rm Im}\mu^2=0$, different analytic expressions have to be used
to compute the rescaled diagonal element of the Green's function $H(Z)$
in these two regions, see Eq.~\eqref{eq:Izfull}. }
\end{figure}

The diagonal elements of the disorder-free lattice Green's function $G_0$
turn out to be independent of the site $i$
\begin{equation}
I(z) =	\langle i| G_0(z) |i\rangle = \int_{\mathcal{B}}
\frac{d\vec{k}}{\Omega} \ \frac{z}{z^2-J^2|f(\vec{k})|^2},
\end{equation}
with $\Omega = 8\pi^2/(3\sqrt{3}a^2)$ the area of the Brillouin zone. We use
the rescaling $\sqrt{3}k_xa/2 \to \alpha$ and $3k_ya/2 \to \beta$. Defining
$Z=z/J$ and $I(z)=H(Z)/J$, little algebra gives
\begin{equation}
\label{eq:H(Z)}
H(Z) = \iint_0^{\pi}\frac{d\alpha d\beta}{\pi^2} \
\frac{Z}{Z^2-(1+4\cos^2\alpha+4\cos \alpha\cos \beta)},
\end{equation}
where $Z$ can be any point in the complex plane except the real interval
$[-3,3]$.

The idea is now to compute \eqref{eq:H(Z)} for real $Z$ outside this interval
and do an analytic continuation in the complex plane. In fact, direct
inspection shows that ${\rm Re}H$ is even in ${\rm Re}Z$ and in ${\rm Im}Z$,
while ${\rm Im}H$ is odd in ${\rm Re}Z$ and in ${\rm Im}Z$. This means that
it is sufficient to compute $H(Z)$ is the first quadrant (${\rm Re}Z \geq 0$,
${\rm Im}Z \geq 0$) of the complex plane and use these parity properties to
infer $H(Z)$ in the other quadrants. By demanding $I(Z)$ to vary smoothly as
$Z$ varies in the complex plane and noticing how $\mu^2$ (see below) crosses
the branch cut of the elliptic integral, we get \cite{morita1971,horiguchi1972}

\begin{equation}
	H(Z)= \left\{\begin{array}{ll}
				\mri\Gamma \, g_0(\Gamma)K( \mu^2_0)
				\hspace{0.4cm} \mbox{for } Z=\mri\Gamma,
				\Gamma>0\\
				\\
				Z \, g(Z)\Bigl(K(\mu^2) + 2\mri K(
				1-\mu^2)\Bigr)\\
				\mbox{for }{\rm Im}\mu^2\leq
				0\;\mbox{and}\;0<{\rm Re}Z\leq1,\\
				\\
				Z \, g(Z)K(\mu^2) \hspace{0.5cm}
				\mbox{otherwise},
				\end{array}\right.\label{eq:Izfull}
\end{equation}
where
\begin{subequations}
\begin{align}
	&g(Z)=\frac{2}{\pi(Z-1)^{\frac{3}{2}}(Z+3)^{\frac{1}{2}}},\\
	&\mu^2 =\frac{16Z}{(Z-1)^3(Z+3)}, \\
	&g_0(\Gamma) =
	\frac{-2}{\pi(\Gamma^2+1)^{\frac{3}{4}}(\Gamma^2+9)^{\frac{1}{4}}},\\
	&\mu^2_0 =
	\frac{16-\left(\sqrt{(\Gamma^2+9)(\Gamma^2+1)}-(\Gamma^2+1)\right)^2}{4(\Gamma^2+1)^{\frac{3}{2}}(\Gamma^2+9)^{\frac{1}{2}}}
\end{align}
\end{subequations}
and
\begin{equation}
	K(\rho^2)=\int_0^{\frac{\pi}{2}}\frac{d\theta}{\sqrt{1-\rho^2\sin^2\theta}}
\end{equation}
is the complete elliptic integral of the first
kind~\cite{gradshteyn2007}. Figure \ref{fig:scba_m2_boundary} gives the
boundary ${\rm Im}\mu^2=0$ in the first quadrant of the complex-$Z$ plane.

\section{Self-consistent Born approximation}
\label{AppendixSCBA}

\subsection{\label{sec:vanHove} Van Hove singularities}
With the parametrization $\Sigma = \gamma J \Exp{-\mri\theta}$, one expects
$\gamma \ll 1$ in the weak disorder regime. A small-parameter expansion in
Eqs.~\eqref{eq:scba1} and \eqref{eq:Izfull}, gives at lowest order $\theta
\approx \pi/2$ and
\begin{equation}
\frac{4}{\gamma} \, \ln\left(\frac{4}{\gamma}\right) = \frac{64\pi J^2}{W^2}
\label{eq:requation}
\end{equation}
Thus,
\begin{equation}
\label{eq:sigmavH}
	\gamma \approx \frac{W^2}{16\pi J^2}\Omega\left(\frac{64\pi
	J^2}{W^2}\right).
\end{equation}
in terms of the Lambert $W$-function $\Omega(\alpha)$ defined for any complex
number $\alpha$ through the identity
\begin{equation}
	\alpha=\Omega(\alpha)\Exp{\Omega(\alpha)}.
\end{equation}
At weak disorder, the asymptotic form for $|\alpha|\gg1$~\cite{corless1996}
\begin{equation}
	\Omega(\alpha)=\ln\alpha-\ln\ln\alpha +
\mathcal{O}\left(\frac{\ln\ln\alpha}{\ln\alpha}\right)\label{eq:productlog}
\end{equation}
yields Eq.~\eqref{eq:vHgam}.

The self-energy in fact obeys the more general relation
\begin{align}
 \Sigma_{{\rm SCBA}}=\frac{4J}{\rho +\mri u}\Omega\big(\Exp{-\mri
2\pi/3}(\rho +\mri u)\big),
\end{align}
where $\rho=(1-\mri 4\pi/3)$ and $u=64\pi J^2/W^2$.
Assuming $|\rho| \ll u$, we get from the latter equation $\theta = \pi/2 +
\delta\theta$ with
\begin{equation}
\delta\theta \approx \frac{\pi}{6\big(\ln(\frac{64\pi
J^2}{W^2})-\ln\ln(\frac{64\pi J^2}{W^2})\big)}.
\end{equation}

\subsection{Band edges}
 We consider the dimensionless ``detuning"  $\delta= 3-E/J$ from the
 band edge at $3J$. We consider $Z=\delta +\Sigma/J \ll 1$ as a small
 parameter. Similar to
what has been done in the previous Section, we find that $Z$
approximately obeys the equation
\begin{align}
\label{eq:ZBE}
Z =
\beta \Bigl(\ln(12/Z)-\mri\pi\Bigr)+\delta,
\end{align}
where $\beta=\frac{\sqrt{3}W^2}{48\pi J^2}$. Using the Lambert $W$-function,
we find
\begin{align}
 \Sigma\approx \beta J \,
 \Omega\big(\frac{-12\Exp{\delta/\beta}}{\beta}\big)-\delta J.
\end{align}
The large-argument expansion of the Lambert $W$-function yields
\begin{align}
 \Sigma\approx
\beta
J\left(\ln(\frac{12}{\beta})-\mri\pi-\ln\left[\ln(\frac{12}{\beta})-\mri\pi+\delta/\beta\right]\right).
\end{align}

\subsection{Scattering mean free path}

For $E<0$, we solve Eq.~\eqref{eq:zmomentum}  to obtain the following
approximate solutions:
\begin{align}
	\frac{a}{\ell}=\left\{\begin{array}{ll}
			 \frac{4}{3} \,
			 \frac{\gamma(E)}{\sqrt{1-\frac{2}{3}{\rm
Re}(Z)}}, & |E|\ll J,\\
			 \frac{2}{\sqrt{3}} \, \gamma(E), & E=-J,\\
			 \frac{2}{\sqrt{3\delta}} \, \gamma(E), &
\delta =3+E/J \ll 1.
		      \end{array}\right.\label{eq:SCBA_ls_approx}
\end{align}
We note that the factor $\sqrt{1-\frac{2}{3}{\rm Re}(Z)}$ is needed when the
energy is sufficiently far from the charge neutrality point (where deviations
from the linear dispersion regime come into play) but not too near the van
Hove singularity (where a different approximation applies). Indeed the real
part of the self-energy is then no longer small compared to the energy $E$
itself, at least for the disorder strengths considered. This is the case
at  $E/J=-0.4$.\\

\section{Derivation of the generalized version of the recursive Green's
function method}\label{sec:recG_derive}
Starting with Eq.~\eqref{eq:splitH_N}, the Born series for the Green's
function $G_N$ reads
\begin{subequations}
\begin{align}
	G_N &= G_{N-1} + G_{N-1}(H_{N}^{\rm  hop} + H_N^{\rm
	slice})G_N,\label{eq:born1}\\
	G_N &= G_{N-1} + G_{N}(H_{N}^{\rm  hop} + H_N^{\rm
	slice})G_{N-1},\label{eq:born2}
\end{align}
\end{subequations}
where $H_{N}^{\rm  hop} = H_{N-1,N}^{\rm  hop} + H_{N,N-1}^{\rm  hop}$.
In Eq.~\eqref{eq:born1}, the term $G_{N-1}H_N^{\rm  slice}G_N$ vanishes
since $H_N^{\rm  slice}$ couples only sites within slice $N$ and one
immediately gets Eq.~\eqref{eq:recG1} after sandwiching with bras
and kets of sites not exceeding the $(N-1)$th slice. Furthermore,
since, by definition, $G_{N-1}$ does not couple to sites in slice
$N$, we immediately recover Eq.~\eqref{eq:recG2} by setting $n=N$
in Eq.~\eqref{eq:recG1}. Eq.~\eqref{eq:recG3} can be obtained from
Eq.~\eqref{eq:born2} through a similar procedure.

To obtain Eq.~\eqref{eq:recG4}, we need to go back to the definition of the
Green's function $(E-H_N)G_N =\mathbb{1}$, which implies
\begin{equation}
	(E-H_{N-1}-H_{N}^{\rm  hop} - H_N^{\rm slice})G_N =\mathbb{1}.
	\label{eq:greendef}
\end{equation}
Sandwiching Eq.~\eqref{eq:greendef} with bras and kets of all sites within
slice $N$, we get
\begin{equation}
	EG^{(N)}_{N,N}-H_{N,N-1}^{\rm  hop}G^{(N)}_{N-1,N}- H_N^{\rm
	slice}G^{(N)}_{N,N}=\mathbb{1}.
	\label{eq:greendef2}
\end{equation}
As Eq.~\eqref{eq:recG2} yields
\begin{equation}
	G^{(N)}_{N-1,N} = G^{(N-1)}_{N-1,N-1}H_{N-1,N}^{\rm  hop}G^{(N)}_{N,N},
\end{equation}
substitution into Eq.~\eqref{eq:greendef2} immediately gives
Eq.~\eqref{eq:recG4}.

\bibliography{disorderV11}
\end{document}